\documentclass[superscriptaddress,prx,twocolumn,nofootinbib]{revtex4-1}
\usepackage{graphicx}
\usepackage{amsmath}
\usepackage{amssymb}
\usepackage{comment}
\usepackage{placeins}
\usepackage[caption=false]{subfig}
\usepackage[colorlinks]{hyperref}
\usepackage{hypcap}

\newcommand{\eq}[1]{Eq.~\hyperref[eq:#1]{(\ref*{eq:#1})}}
\renewcommand{\sec}[1]{\hyperref[sec:#1]{Section~\ref*{sec:#1}}}
\newcommand{\app}[1]{\hyperref[app:#1]{Appendix~\ref*{app:#1}}}
\newcommand{\tab}[1]{\hyperref[tab:#1]{Table~\ref*{tab:#1}}}
\newcommand{\fig}[1]{\hyperref[fig:#1]{Figure~\ref*{fig:#1}}}
\newcommand{\figa}[2]{\hyperref[fig:#1]{Figure~\ref*{fig:#1}#2}}
\newcommand{\figx}[2]{\hyperref[fig:#1]{Figure~\ref*{fig:#1}(#2)}}
\newcommand{\thm}[1]{\hyperref[thm:#1]{Theorem~\ref*{thm:#1}}}
\newcommand{\lem}[1]{\hyperref[lem:#1]{Lemma~\ref*{lem:#1}}}
\newcommand{\cor}[1]{\hyperref[cor:#1]{Corollary~\ref*{cor:#1}}}
\newcommand{\defn}[1]{\hyperref[def:#1]{Definition~\ref*{def:#1}}}
\newcommand{\alg}[1]{\hyperref[alg:#1]{Algorithm~\ref*{alg:#1}}}

\def\bra#1{\mathinner{\langle{#1}|}}
\def\ket#1{\mathinner{|{#1}\rangle}}
\newcommand{\braket}[2]{\langle #1|#2\rangle}

\newcommand{\ignore}[1]{}

\newcommand{\be}{\begin{equation}}
\newcommand{\ee}{\end{equation}}
\newcommand{\ba}{\begin{eqnarray}}
\newcommand{\ea}{\end{eqnarray}}

\input{Qcircuit}
\begin{document}

\title{Chemical Basis of Trotter-Suzuki Errors in Quantum Chemistry Simulation}

\date{\today}

\author{Ryan Babbush} 
\affiliation{Quantum Architectures and Computation Group, Microsoft Research, Redmond, WA 98052}

\author{Jarrod McClean} 
\affiliation{Department of Chemistry and Chemical Biology, Harvard University, Cambridge, MA 02138}

\author{Dave Wecker}
\affiliation{Quantum Architectures and Computation Group, Microsoft Research, Redmond, WA 98052}

\author{Al\'an Aspuru-Guzik}
\affiliation{Department of Chemistry and Chemical Biology, Harvard University, Cambridge, MA 02138}

\author{Nathan Wiebe}
\affiliation{Quantum Architectures and Computation Group, Microsoft Research, Redmond, WA 98052}

\begin{abstract}

Although the simulation of quantum chemistry is one of the most anticipated applications of quantum computing, the scaling of known upper bounds on the complexity of these algorithms is daunting. Prior work has bounded errors due to Trotterization in terms of the norm of the error operator and analyzed scaling with respect to the number of spin orbitals. However, we find that these error bounds can be loose by up to sixteen orders of magnitude for some molecules. Furthermore, numerical results for small systems fail to reveal any clear correlation between ground state error and number of spin orbitals. We instead argue that chemical properties, such as the maximum nuclear charge in a molecule and the filling fraction of orbitals, can be decisive for determining the cost of a quantum simulation. Our analysis motivates several strategies to use classical processing to further reduce the required Trotter step size and to estimate the necessary number of steps, without requiring additional quantum resources. Finally, we demonstrate improved methods for state preparation techniques which are asymptotically superior to proposals in the simulation literature.
\end{abstract}

\maketitle

\section{Introduction}

The idea that the simulation of quantum systems would be efficient on a quantum computer dates back to Feynman's original work on quantum mechanical computers \cite{Feynman1982}. Almost a decade after Abrams and Lloyd  \cite{Abrams1997} demonstrated a scalable scheme for the quantum simulation of fermions, Aspuru-Guzik \emph{et al.} \cite{Aspuru-Guzik2006} proposed that these techniques could be used to efficiently determine the ground state energy of molecular Hamiltonians, solving what chemists refer to as the electronic structure problem. Since then, a great deal of work has focused on specific strategies for the quantum simulation of quantum chemistry. While most of these approaches are based on a second quantized representation of the problem making use of both phase estimation and Trotterization \cite{Ortiz2001,Aspuru-Guzik2006,Wang2008,Wang2009,Whitfield2010,CodyJones2012,Seeley2012,Wecker2013,Hastings2014,Poulin2014,McClean2014}, recently some have proposed alternative schemes such as the quantum variational eigensolver \cite{McClean2013}, an adiabiatic algorithm \cite{Babbush2014} and an oracular approach based on a 1-sparse decomposition of the configuration interaction Hamiltonian \cite{Toloui2013}. In fact, quantum chemistry is such a popular application that toy problems in chemistry have been solved on a variety of experimental quantum information processors which include quantum optical systems \cite{Lanyon2009,McClean2013}, nuclear magnetic resonance \cite{Du2010,Lu2011} and solid-state Nitrogen-vacancy center systems \cite{Wang2014}.

Recently, a series of papers \cite{Wecker2013,Hastings2014,Poulin2014,McClean2014} has provided improved analytical and empirical bounds on the resources required to simulate classically intractable benchmarks using a quantum computer. While the initial findings in \cite{Wecker2013} were pessimistic, improvements in both bounds and algorithms introduced in \cite{Hastings2014} and \cite{Poulin2014} have reduced these estimates by more than thirteen orders of magnitude for simulations of Ferredoxin. The primary contribution of \cite{McClean2014} was to point out that in the limit of large molecules, the use of a local basis can substantially reduce asymptotic complexity of these algorithms. In this paper we build on the findings of  \cite{Wecker2013,Hastings2014,Poulin2014,McClean2014} to offer new perspectives regarding the scaling of the second quantized, Trotterized, phase estimation algorithm for quantum chemistry. In particular, we question a basic assumption implicit in all of these works: that the Trotter error explicitly depends on the number of spin orbitals being simulated.

Instead, we argue that chemical properties such as the filling fraction of electrons in a given basis, the particular choice of orbital basis and the nuclear potential play a more significant role in determining the Trotter error than does the number of spin orbitals for small molecules. We support these arguments with numerical analysis based on the explicit computation of the Trotter error operator derived in \cite{Poulin2014}. Additionally, we show that classically tractable approximations to the ground state wavefunction can be used to efficiently estimate the Trotter error expected in a particular ground state simulation. This result is of significant practical importance because without a procedure for estimating the Trotter error, one must rely on analytical error bounds which (as we show) tend to overestimate the ground state error by many orders of magnitude. Finally, we show asymptotically improved circuits for state preparation based on these classical ansatz states.

\subsection{The electronic structure problem}

The electronic structure problem is to estimate the energy of electrons interacting in a fixed nuclear potential to within an additive error of $\epsilon$. This Hamiltonian may be written as,
\begin{align}
\label{eq:electronic}
 H = - \sum_i \frac{\nabla_{r_i}^2}{2} - \sum_{i,j} \frac{Z_i}{|R_i - r_j|} + \sum_{i, j>i} \frac{1}{|r_i - r_j|}
\end{align}
where we have used atomic units, $\{R_i\}$ denotes nuclear coordinates, $\{r_i\}$ electronic coordinates, and $\{Z_i\}$ nuclear charge. Often times, the utility of these energies is to provide Born-Oppenheimer surfaces for molecular modeling at finite temperatures. Usually, chemists are interested in obtaining free energy landscapes which provide mechanistic insight into chemical events of significant practical importance such as drug binding, catalysis and material properties. These free energy landscapes must be extremely accurate as chemical rates are exponentially sensitive to changes in free energy. Under typical laboratory conditions of room temperature and atmospheric pressure, ``chemical accuracy'' is required which sets $\epsilon$ to the order of $10^{-3}$ hartree \cite{Helgaker2013} where $1$ hartree is $\frac{\hbar^2}{m_e e^2 a_0^2}$ and $m_e$, $e$ and $a_0$ denote the mass of an electron, charge of an electron and Bohr radius, respectively.

We represent the electronic structure Hamiltonian in second quantization \cite{Helgaker2013} as this requires significantly fewer qubits than approaches using the first quantized Hamiltonian \cite{Zalka1998, Kassal2008},
\begin{align}
 H = \sum_{pq} h_{pq} a_p^{\dagger}a_q + \frac{1}{2} \sum_{pqrs} h_{pqrs} a_p^{\dagger} a_q^{\dagger} a_r a_s
\label{eq:2nd}
\end{align}
in which creation and annihilation operators act on a basis of orthogonal spin orbitals, $\{\varphi_i\}$ and the one-electron and two-electron integrals are
\begin{align}
 &h_{pq} = \int d\sigma\ \varphi_p^*(\sigma) \left(-\frac{\nabla_{r}^2}{2} - \sum_{i} \frac{Z_i}{|R_i - r|} \right)\varphi_q(\sigma) \label{eq:single_int}\\
 &h_{pqrs} = \int d\sigma_1\ d\sigma_2\ \frac{ \varphi_p^*(\sigma_1) \varphi_q^*(\sigma_2)  \varphi_s(\sigma_1) \varphi_r(\sigma_2) }{|r_1 - r_2|} \label{eq:double_int}
\end{align}
where $\sigma_i$ contains spatial and spin degrees of freedom for the electrons. The operators $a_p^\dagger$ and $a_r$ obey the fermionic anti-commutation relations
\begin{align}
\label{eq:anticomm}
 \{a_p^\dagger, a_r\} = \delta_{p,r}, \quad \quad \{a_p^\dagger, a_r^\dagger\} = \{a_p, a_r\} = 0.
\end{align}

In principle, the number of spin orbitals used to represent a molecule is not a property of the molecule. However, the quantum chemistry community has certain conventions (based on periodic trends) for the number of spin orbitals that should be used for each atom in the period table, depending on the desired level accuracy in the calculation. In a minimal basis, first period atoms receive two spin orbitals, second period atoms receive 10 spin orbitals and third period atoms receive 18 spin orbitals. The reasoning behind this scheme is that the most important orbitals are those which have a principal quantum number less than or equal to that of the highest occupied orbital according to Hund's rules.

In addition to choosing a spatial basis, one must choose an orbital basis that associates orthogonal spatial functions constructed from the spatial basis, with the second quantized sites. Throughout this paper we investigate three such orbital basis sets: the ``local basis'' is the set of orthogonal atomic orbitals discussed in \cite{McClean2014}, the ``canonical basis'' is the Hartree-Fock molecular orbitals, and the ``natural basis'' is that which diagonalizes the one-electron density matrices associated with the exact ground state\footnote{The natural basis can be well approximated without performing an exact calculation by repeating truncated configuration interaction calculations from reference states defined using the natural orbitals associated with a previous solution.}. It is worth pointing out that the canonical orbitals are the natural orbitals of a Hartree-Fock calculation using a single determinant.

From~\eq{2nd}, we see that the number of terms in the Hamiltonian scales as $\Theta\left(N^4\right)$\footnote{We use the typical computer science convention that $f\in \Theta(g)$, for any functions $f$ and $g$,  if $f$ is asymptotically upper and lower bounded by a multiple of $g$, ${\cal O}$ indicates an asymptotic upper bound, $\tilde{{\cal O}}$ indicates an asymptotic upper bound up to polylogarithmic factors, $\Omega$ indicates the asymptotic lower bound and $f\in o(g)$ implies $f/g\rightarrow 0$ in the asymptotic limit.}. However, McClean \emph{et al.} \cite{McClean2014} recently pointed out that the basis functions decay super-exponentially with distance in a local basis.  This means that the integrals in~\eq{single_int} and~\eq{double_int} will be negligibly small for many of the orbitals which in turn allows the number of terms in the Hamiltonian to be truncated to $\tilde{\mathcal O}\left(N^2\right)$ or $\tilde{\mathcal O}\left(N\right)$ depending on the size and geometry of the molecule. All of the particular benchmarks studied in this paper involve less than four atoms and so we consider the number of non-negligible terms in the Hamiltonian to scale as $\Theta(N^4)$, even in a local basis.

\subsection{Quantum simulation of quantum chemistry}

The electronic structure problem is classically intractable to current methods even after discretizing the Hilbert space.  This intractability can be understood as a consequence of the exponential size of the Hilbert space for the second quantized Hamiltonian.  Similarly, existing methods such as configuration interaction, require consideration of a number of electronic configuration states that increases exponentially as the approximation becomes more exact.  Quantum simulation offers a way to circumvent these challenges by directly mapping the chemical system onto a set of qubits that can be manipulated using a quantum computer. The particular problem that we focus on is the problem of computing the ground state energy of the system.  Other important physical quantities such as dipole moments can be found by evaluating their expectation value with respect to the prepared state. The simulation problem that we consider is as follows.\\

\textbf{Problem:} \emph{Assume that the user is provided with a classical database containing $h_{pq}$ and $h_{pqrs}$ for a molecule with $N$ spin orbitals and a blackbox state preparation algorithm that prepares an approximation $\ket{\tilde{0}}$ to the ground state $\ket{0}$ such that $|\braket{\tilde{0}}{0}|^2 \in \Omega \left(\textrm{poly}\left(N^{-1}\right)\right)$.  Design a quantum circuit that uses these elements to estimate the ground state energy of~\eq{2nd} within additive error $\epsilon$ using  a minimal expected number of gates and qubits.}\newpage

Most proposals for quantum computer simulation of chemical systems use similar strategies to solve this problem.  The first step involves translating the basis of the second quantized Hamiltonian to that of the quantum computer.  The standard way to do this is to use the occupation number basis in which individual qubits encode the occupation of a spin orbital.  For example, the state $\ket{00011}$ would refer to an electronic state where the first two spin orbitals are occupied and the remaining three spin orbitals are unoccupied.

Although representing states is trivial, representing the Hamiltonian is not.  The reason is that, although it may seem that the creation and annihilation operators $a_i^\dagger$ and $a_i$ are translated to $(X_i-iY_i)/{2}$ and $(X_i+iY_i)/{2}$ respectively, the resulting operators do not obey the anti-commutation relations in~\eq{anticomm}.  This problem is addressed by using either the Jordan-Wigner transformation \cite{Somma2002,Aspuru-Guzik2006} or the Bravyi-Kitaev transformation \cite{Bravyi2000,Seeley2012} to modify these operators to have the correct anti-commutation relations. Importantly, the operators that result from using either of these representations are tensor products of Pauli operations.  While the number of such terms in the transformed Hamiltonian scales as ${\cal O}(N^4)$ using both approaches, the locality (i.e. many-body order) of these terms scales as ${\cal O}\left(N\right)$ under the Jordan-Wigner transformation and ${\cal O}\left(\log N\right)$ under the Bravyi-Kitaev transformation \cite{Seeley2012}.

Since exponentials of a polynomial number of Pauli operators are known to be efficiently simulatable, $e^{-iHt}\ket{\tilde{\psi}}$ can be implemented using a polynomial number of gates using a quantum computer.  There are many different approaches that can be used to achieve this and the majority of these rely on Trotter decompositions, which we will discuss in more detail later.  However, each of these methods solves a dynamical simulation problem and does not directly solve the ground state energy estimation problem.  The phase estimation algorithm provides the connection needed to relate the eigenvalue estimation problem to a dynamical simulation problem.

The quantum phase estimation algorithm (PEA) uses a quantum computer to efficiently estimate energies from the phases $\{\theta_n\left(t\right)\}$ accumulated during time evolution under a propagator $U_H\left(t\right)$ associated with the Hamiltonian of interest $H$; i.e.
\begin{align}
& e^{i H t} \ket{n} = U_H\left(t\right)\ket{\psi_n} = e^{i \theta_n\left(t\right)} \ket{\psi_n}\\
& \theta_n\left(t\right) = \left(E_n t\right) \mod 2 \pi
 \end{align}
 where $\{\ket{\psi_n}\}$ and $\{E_n\}$ represent eigenstates and eigenvalues of $H$. If we initialize a quantum register in a state $\ket{\tilde{\psi_0}}$ then time evolution under a static Hamiltonian produces the superposition,
\begin{align}
U_H\left(t\right) \ket{\tilde{\psi_0}} = \left(\sum_{n = 0}^{2^N - 1} e^{i \theta_n\left(t\right)} \ket{\psi_n}\!\!\bra{\psi_n}\right)\ket{\tilde{\psi_0}}.
\label{super}
\end{align}
Measuring the phase of this superposition projects the system to state $\ket{\psi_0}$ with probability $|\braket{\psi_0}{\tilde{\psi_0}}|^2$. Thus, under the assumptions of our problem, at most a polynomial number of repetitions of the phase estimation algorithm will be needed to find the ground state energy.

There are obviously two contributions to the cost of solving the electronic structure problem via quantum computing:  ($a$) the overlap $|\braket{\psi_0}{\tilde{\psi_0}}|^2$ and (b) the cost of simulating the dynamics of the system.  Since the overlap is independent of the simulation method used (to second-order in perturbation theory) most work on the topic has focused on reducing the latter cost.  We discuss both of these issues in the following.

Our main focus is on Trotter-Suzuki based methods, which involve a discretization of the time evolution known as Trotterization.  Trotterization approximates $U_H\left(t\right)$ as a series of time steps known as ``Trotter slices'' during which only one of the Hamiltonian terms is actually active.  A Trotter series containing $\mu$ Trotter slices is said to have a ``Trotter number'' of $\mu$ and the error in this approximation, which arises from non-commutativity of the Hamiltonian terms, vanishes as $\mu\rightarrow \infty$. For a fixed order Trotter-Suzuki formula, each Trotter slice contains a number of gates that is proportional to the number of terms in the Hamiltonian,  $m$.  The value of $m$ depends on basis and molecular size and its scaling with $N$ ranges from $\tilde{\mathcal{O}}\left(N\right) -\tilde{\mathcal{O}}\left(N^4\right)$. Since the the total complexity of the quantum simulation circuit for chemistry is $\tilde{\mathcal{O}}\left(m\mu \right)$, understanding how both of these terms scales is vital for determining whether quantum chemistry will be viable on small scale quantum computers.

\begin{figure}[t!]
\centering
\includegraphics[width=8cm]{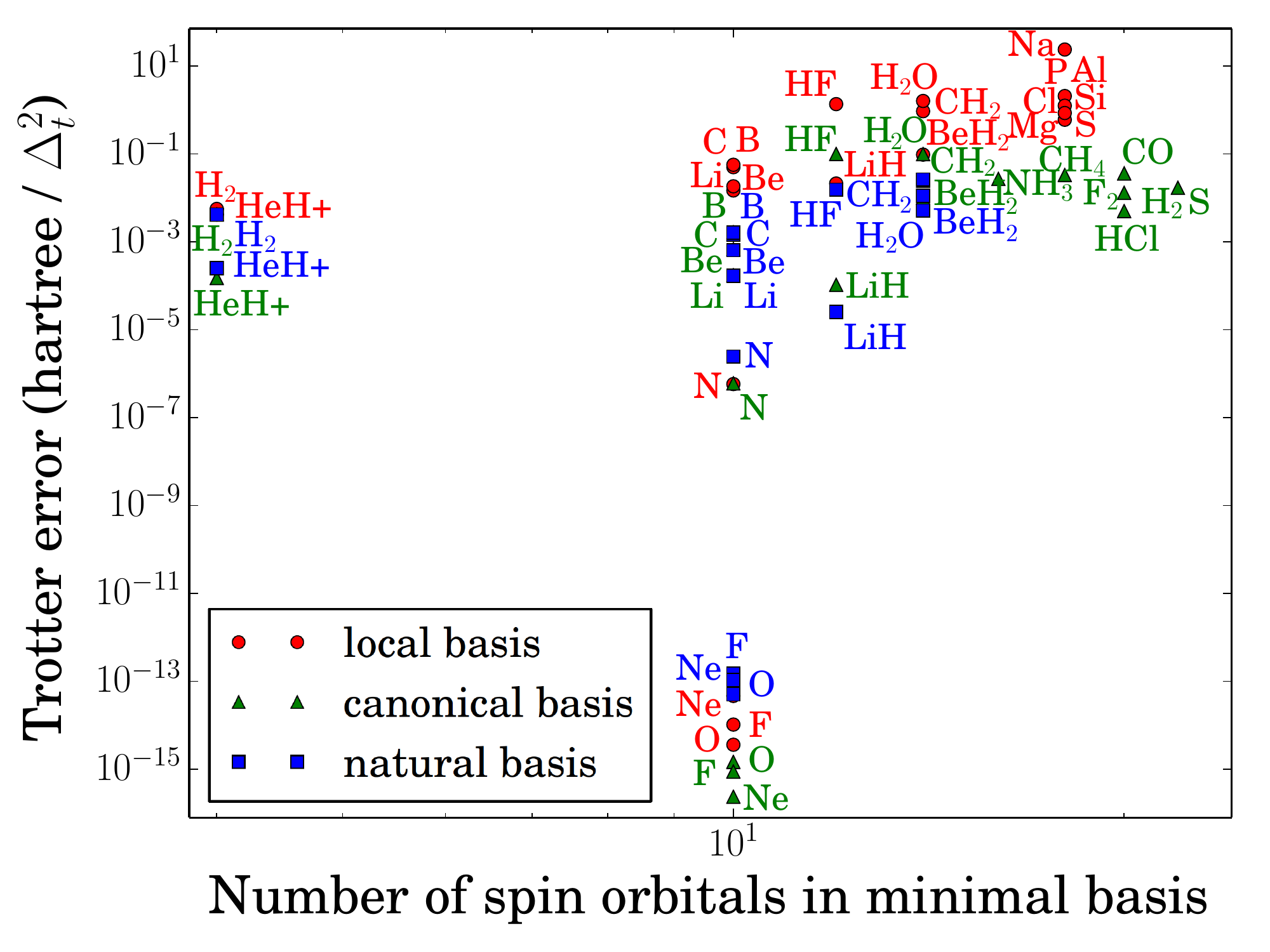}
\caption{Spin-orbitals versus ground state Trotter error for various molecular benchmarks in three different basis sets. Despite analytical predictions to the contrary (in prior works), it would appear that no clear relation holds between the Trotter error induced on the ground state and the number of spin orbitals for these benchmarks.}
\label{fig:orbitals_error}
\end{figure}

The big question that several recent papers have attempted to address is: ``how does $\mu$ scale with $N$?''  Indeed, this issue is central to the optimizations introduced in many of these simulation methods. 
Given the importance of this issue in the literature, the data in \fig{orbitals_error} may come as a complete surprise.  We see there that for modestly small molecules, the error in the second-order Trotter-Suzuki formula does not have a clear functional dependence on $N$.  This is especially surprising for cases of canonical and natural orbitals where there is  little evidence of even an increasing trend in the error as a function of $N$.  This lack of monotonicity is particularly striking for the atoms N, O, F, Ne which show negligibly small Trotter errors.  In fact, for these molecules (along with others such as Helium Hydride and Lithium Hydride) $\mu=1$ or $\mu=2$ is sufficient to achieve chemical accuracy despite the fact that their Hamiltonians contain hundreds of non-commuting terms.  

In order to understand why the Trotter error deviates so strongly from prior expectations, we analyze a leading order perturbative expression for the error in the second-order Trotter formula.  The insights gained from this analysis raise an interesting point: although there is not a strong correlation between $N$ and the Trotter error, other chemical properties play a decisive role in the Trotter error.  This forces us to reconsider how we conceptualize the scaling of quantum chemistry simulation relative to prior results in quantum simulation, e.g. \cite{Abrams1997,Aspuru-Guzik2006,Kassal2008,Aharonov2003,Childs2013a,Berry2012,Berry2013,Hastings2014,Poulin2014,McClean2014}.

\section{Analysis of Trotter error operator}

The second-order Trotter-Suzuki decomposition allows us to approximate the propagator as a series of unitaries corresponding to the individual Hamiltonian terms. In particular, the second-order\footnote{Note that in work that focuses on high-order Trotter-Suzuki formulas \eq{trotter} is often called the first-order Trotter Suzuki formula because it is the lowest iteration order in Suzuki's iterative construction of high-order splitting formulas.}
 Trotter formula gives us,
\begin{align}
\label{eq:trotter}
U_H^{\textrm{TS}}\left(\Delta_t\right)\equiv \prod_{\alpha=0}^{m - 1} U_{m - \alpha}\left(\frac{\Delta_t}{2}\right)\prod_{\alpha=1}^{m} U_\alpha\left(\frac{\Delta_t}{2}\right)
\end{align}
where,
\begin{align}
U_\alpha\left(\frac{\Delta_t}{2}\right) = e^{-i H_\alpha \Delta_t / 2}.
\end{align}
The second-order formula applies each unitary twice with the second half of the Trotter series in reverse order of the first half to cancel out error terms in the ground state energy that would arise at first-order in $\Delta_t$. We use this to make the approximation, valid for sufficiently small values of $\Delta_t$, that
\begin{align}
U = e^{i H t} \approx \left(U^\textrm{TS}\left(\Delta_t\right)\right)^\mu, \quad \Delta_t = t/\mu.
\end{align}
Poulin \emph{et al.} \cite{Poulin2014} focus on bounding the error in this approximation with the Baker-Campbell-Hausdorff (BCH) formula,
\begin{align}
\label{eq:bch}
\log\left(e^X e^Y\right) & = X + Y + \frac{1}{2}\left[X,Y\right]\\
&  + \frac{1}{12}\left[X, \left[X, Y\right]\right] - \frac{1}{12}\left[Y, \left[X,Y\right]\right] + ...\nonumber
\end{align}
By recursively applying \eq{bch} to \eq{trotter}, the error operator may be written as $V=\sum_{j=1}^\infty V^{(j)}$.  The leading order term in this expansion is,
\begin{align}
\label{eq:error_op}
V^{\left(1\right)} = -\frac{\Delta_t^2}{12}\sum_{\alpha \leq \beta} \sum_{\beta} \sum_{\gamma < \beta}\left[H_\alpha\left(1 - \frac{\delta_{\alpha, \beta}}{2}\right), \left[H_\beta, H_\gamma\right] \right]
\end{align}
with errors on the order of ${\cal O}\left(\Delta_t^4\right)$. 

The leading order shift in the energy of the $i^\textrm{th}$ eigenstate is given by non-degenerate perturbation theory as,
\begin{align}
\label{eq:shift}
\Delta E_i = \bra{\psi_i} V^{\left(1\right)} \ket{\psi_i} + {\cal O}\left(\Delta_t^4\right)
\end{align}
where $H \ket{\psi_i} = E_i \ket{\psi_i}$. Solving the electronic structure problem requires fixed precision in the energy, i.e. $\Delta E = {\cal O} \left(1\right)$. This suggests that we must shrink the time step for larger problem instances in order to offset any increase in Trotter error. In order to make the leading order shift in the energy eigenvalue at most $\delta$ it suffices to take
\begin{align}
\mu\! &=\! {\cal O}\!\left(\! t\sqrt{\frac{1}{\delta}\!\left \langle \sum_{\alpha \leq \beta} \sum_{\beta} \sum_{\gamma < \beta}\left[H_\alpha\!\left(1 \!-\! \frac{\delta_{\alpha, \beta}}{2}\right)\!, \left[H_\beta, H_\gamma\right] \right] \!\right\rangle}\!\right).\label{eq:mueqn}
\end{align}
Higher-order Trotter-Suzuki algorithms can be used to reduce the scaling of $\mu$; however they require a number of gates that scales exponentially with the order of the Trotter formula.  This means that for many problems with modest error tolerances, the second-order Trotter formula~\eq{trotter} yields the most efficient results.  Although a similar expression based on degenerate perturbation theory must be used for molecules near disassociation, in most practical cases~\eq{mueqn} will accurately predict the required Trotter number in the limit of small $\delta$.

In practice, it is difficult to determine precisely how this error scales with problem size for real molecules. By inspection of \eq{error_op}, a loose bound of $\mu= {\cal O}\left(N^{5}\right)$ is obtained \cite{Poulin2014}. This bound is obtained by recognizing that the double commutator sum in \eq{error_op} contains $\mathcal{O}(N^{12})$ terms but only $\mathcal{O}(N^{10})$ such terms are non--zero.  In some cases, such as large molecules represented in a local orbital basis, many of these interactions can be neglected and the actual scaling of $\mu$ needed to achieve chemical accuracy may be closer to $\mu = \tilde{\mathcal{O}}\left(N^3 \right)$ or $\mu = \tilde{\mathcal{O}}\left( N^{3/2}\right)$.

All of these scalings follow from worst case assumptions about the error and liberal application of the triangle inequality.  Such arguments are not sufficient to explain the data in~\fig{orbitals_error} which does not show a clear dependence of $\mu$ on $N$.  We therefore focus in the remainder on two quantities: ($a$) the error in the ground state energy and ($b$) the operator norm of the Trotter error operator.  While ($a$) is the best measure of the error in quantum chemistry simulation, we also focus on $(b)$ because it upper bounds ($a$) and because it can be well approximated without diagonalizing the Hamiltonian.

In the numerics that follow we construct error operators by explicitly computing all ${\cal O}(N^{10})$ nonzero terms in~\eq{error_op}. Once all the terms in the error operator are constructed, we simplify the resulting expression by normal-ordering the result.  Here normal-ordering refers to a sorting process where any chain of creation and annihilation operators that result from~\eq{error_op} are reordered such that creation operators always occur at the left-most part of the chain.  This reordering is done by using the anti-commutation relations in~\eq{anticomm}.  For example, $a_2 a_1 a^\dagger_1 a_3^\dagger=a_1^\dagger a_3^\dagger a_1a_2-a_3^\dagger a_2$.  These normal-ordered terms are then grouped, allowing their actions on computational basis states to be efficiently computed.

The Trotter scheme we investigate does not use the coalescing strategies introduced in \cite{Poulin2014}, which would surely lead to even more error cancellation. We use a minimal spatial basis (STO-6G). The Trotter series is ordered in the ``interleaving'' scheme introduced in \cite{Hastings2014} and PQRS terms are ordered lexicographically. All molecular integrals in this work were calculated at equilibrium configurations using the GAMESS electronic structure package \cite{gamess1,gamess2}. While computing the error operator is efficient, evaluating the error operator on an eigenstate of the Hamiltonian cannot be performed in polynomial time on a classical computer. Due to the expensive nature of these calculations, we limit our investigation to benchmarks containing less than twenty spin orbitals. We study the scaling of the norm of the Trotter error operator as this quantity is the focus of analytical bounds introduced in \cite{Wecker2013} and \cite{Poulin2014}. Though the bounds in  \cite{Poulin2014} are based on a upper bound for the operator norm of the error operator, here we use the exact value of $\|V^{(1)}\|$.

\subsection{Comparison of norm of error operator and ground state error}

An important question to ask is, ``how does the error in the simulated ground state energy compare to that predicted by the norm of the error operator?'' This is important for two reasons.  The first reason is that there can be substantial cancellation in the sum implicit in~\eq{shift}.  This effect is also discussed in~\cite{Poulin2014}.  The second reason is that the ground state may only have limited overlap with the eigenstates of the error operator that have large eigenvalues.  We will discuss these two effects in detail later, but for now it suffices to ask how substantial the differences between the two measures are.
\begin{figure}
\centering
\includegraphics[width=8cm]{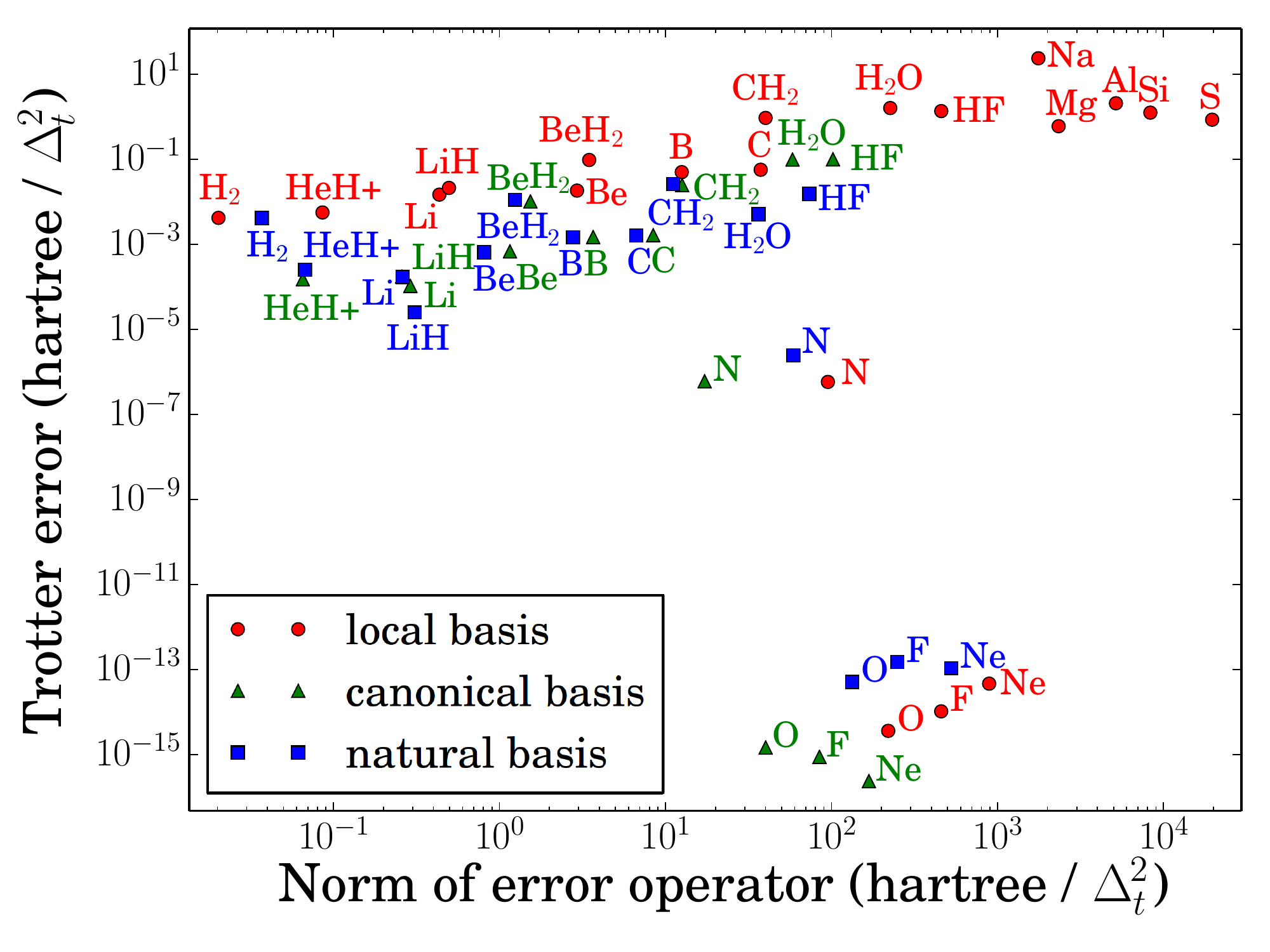}
\caption{A comparison between the norm of the error operator and the error induced in the ground state. Notice that in many cases the basis of natural orbitals have the lowest Trotter error (especially for examples with large Trotter error).}
\label{fig:basis_set_compare}
\vspace{-.25cm}
\end{figure}

\fig{basis_set_compare} shows that substantial differences exist between the computed Trotter error and the norm of the error operator.  In particular, for O, F and Ne these discrepancies can be as large as sixteen orders of magnitude.  Other molecules, such as ${\rm H}_2$O and HF differ by only two orders of magnitude.  This shows that existing estimates of the error can ludicrously overestimate the error in Trotter Suzuki formulas if the properties of the ground state are not also taken into account. Similar comparable results have also been observed for random many-body Hamiltonians \cite{Raeisi2012}.

To see this, let us consider Ne.  By the convention for second-period atoms, Ne is given $10$ spin orbitals in a minimal basis but it also has $10$ electrons.  This means that all of its spin orbitals will be occupied, i.e. $\ket{\psi_0} = \ket{1}^{\otimes 10}$.  If we consider the action of a single normal-ordered term from~\eq{shift}, $\alpha \,a^{\dagger}_{p_1} \cdots a^{\dagger}_{p_5} a_{q_1}\cdots a_{q_5}$, then we see that $\bra{\psi_0} \alpha\, a^{\dagger}_{p_1} \cdots a^{\dagger}_{p_5} a_{q_1}\cdots a_{q_5} \ket{\psi_0}=0$ unless $\{p_1,\ldots,p_5\}=\{q_1,\ldots,q_5\}$ up to permutations.
Thus, the vast majority of the terms present in the error operator will evaluate to zero, irrespective of the magnitude of their coefficients.  A similar argument can be made for F and O except that the ground state will no longer precisely be the Hartree-Fock state and instead will be a linear combination of computational basis states.  Nonetheless, it is easy to see that the vast majority of these expectation values will be zero for these highly constrained systems.  We therefore expect from this argument that molecules that have spin orbitals that are nearly fully occupied will have abnormally low error compared to molecules that are half filled where the dimension of the space is maximal for a given number of basis functions.  This not only justifies the shockingly small error in N, O, F, and Ne but also explains why only considering the norm of the error operator obscures this trend. 

For most benchmarks there is still evidence of correlation between the norm of the error operator and the Trotter error. This means that trends in the norm of the error operator are often reflected in the simulation error.  As we have seen, the properties of the molecules in question can change the nature of this relationship.

\subsection{Dependence on basis}

\begin{center}
\begin{table}
\label{basis_table}
\begin{tabular*}{.925\columnwidth}{@{\extracolsep{\fill}}  c c c c }
\noalign{\vskip 1.5mm}
\hline\hline
\noalign{\vskip 1.5mm}
Basis & Type & Orbitals & error / norm \\
\noalign{\vskip .5mm}
\hline
\noalign{\vskip 1mm}
STO-6G & local & 4 & 0.2063 \\
3-21G & local & 8 & 0.0568 \\
6-31G & local & 8 & 0.0592 \\
6-31++G & local & 12 & 0.0328 \\
\noalign{\vskip .5mm}
\hline
\noalign{\vskip 1mm}
STO-6G & canonical & 4 & 0.1131 \\
3-21G & canonical & 8 & 0.0231 \\
6-31G & canonical & 8 & 0.0242 \\
6-31++G & canonical & 12 & 0.0108 \\
\noalign{\vskip .5mm}
\hline
\noalign{\vskip 1mm}
STO-6G & natural & 4 & 0.1131 \\
3-21G & natural & 8 & 0.0472 \\
6-31G & natural & 8 & 0.0547 \\
6-31++G & natural & 12 & 0.0194\\
\noalign{\vskip .5mm}
\hline\hline
\end{tabular*}
\caption{Ratio of ground state error to error operator norm for molecular hydrogen in various basis sets.}\label{tab:errorratio}
\vspace{-.4cm}
\end{table}
\end{center}
\vspace{-1cm}

In addition to showing that Trotter error in the ground state is usually substantially less than the error operator norm, \fig{basis_set_compare} suggests that the error is also basis dependent. While previous works have focused on the local and canonical basis sets, this figure suggests that using natural orbitals can often lower Trotter error by several orders of magnitude relative to a local orbital basis. 

Furthermore, we argue that the discrepancy between error norm and ground state error increases with the number of spin orbitals to such an extent that the former should not be used to make arguments about the asymptotic scaling of the latter. One can always add more spin orbitals to a molecular Hamiltonian but given a reasonable orbital basis, the ground state and physically meaningful excited states will have increasingly limited occupancy in high energy orbitals. In this context, the energy of an orbital is understood to mean the energy of a single electron occupying that orbital in the absence of other electrons (appropriate for atomic orbitals) or in the presence of the average density of all other electrons (appropriate for the canonical orbitals). Additionally, the natural orbital basis is known to have the property that states with an odd number of excitations from ground state reference often have negligible overlap with the exact ground state \cite{Benavides2014}.

While the error operator will inevitably contain many terms involving excitations to and from these high energy spin orbitals, eigenstates of physical interest (e.g. the ground state) are superpositions of configurations which have a limited number of excitations. Accordingly, terms involving combinations of high energy orbitals are not expected to significantly contribute to the error induced in relevant eigenstates despite increasing the norm of the error operator. This principle is demonstrated in~\tab{errorratio} which shows the ratio between ground state error and error norm for molecular hydrogen in various basis sets.

\subsection{Dependence on nuclear charge}

\begin{figure}
\centering
\includegraphics[width=8cm]{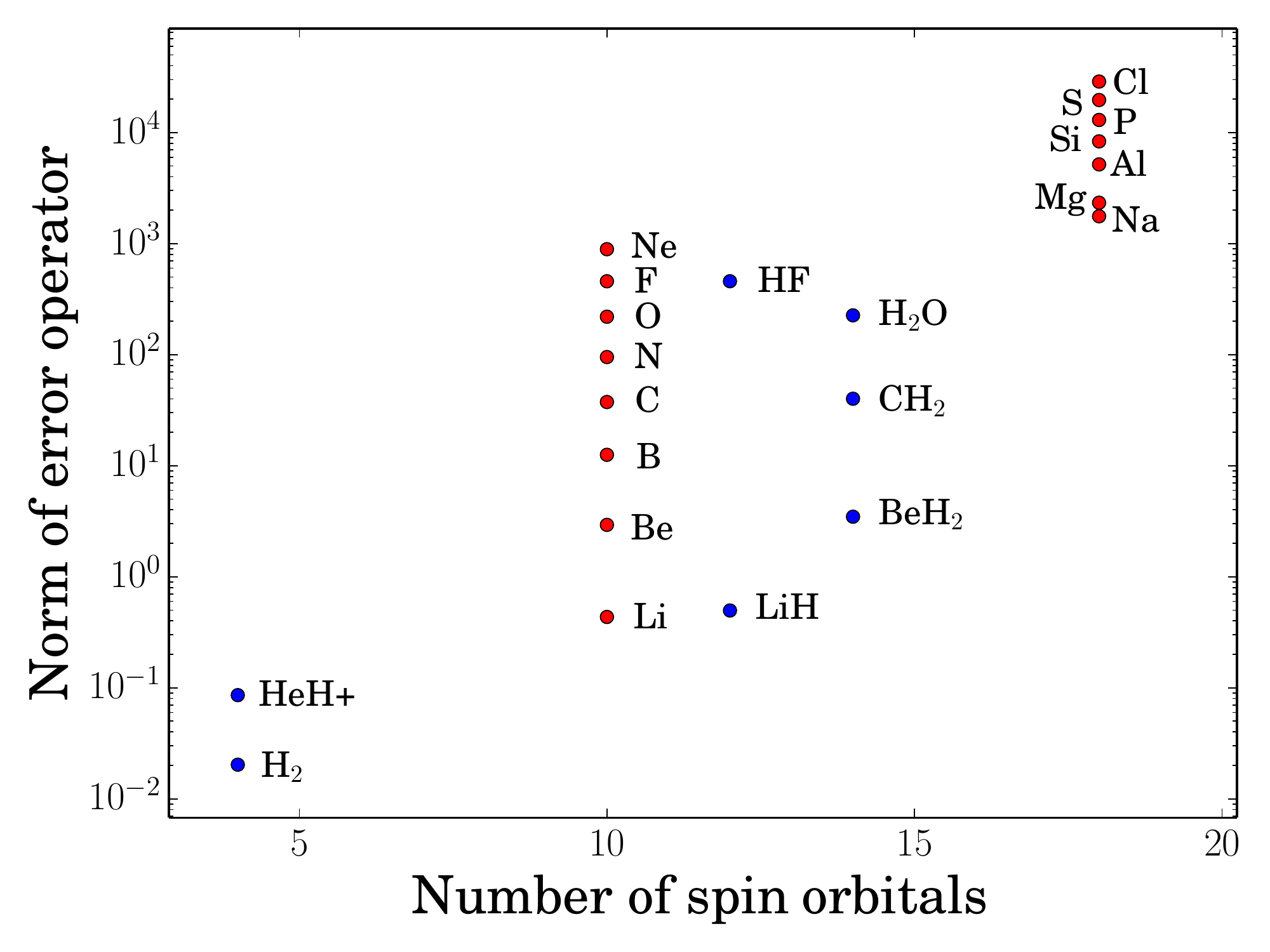}
\caption{We correlate the number of spin orbitals with the norm of the error operator in the local basis. The semblance of a positive slope appears to be a symptom of increasing nuclear charge as the number of spin orbitals increase.  Red dots are atoms and blue dots are molecules.}
\label{fig:orbitals_vs_norm}
\end{figure}

\begin{figure}
\centering
\includegraphics[width=8cm]{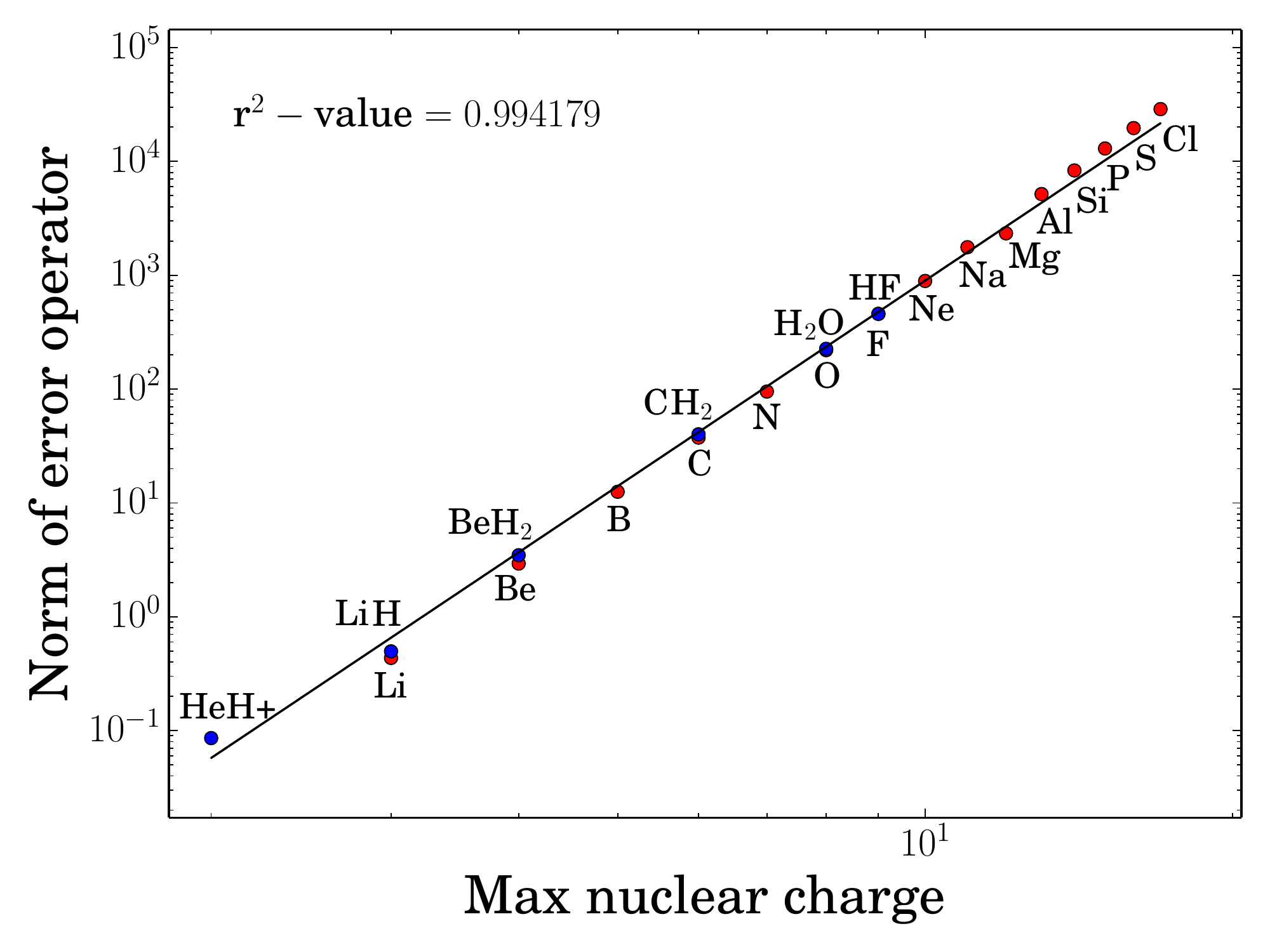}
\caption{The norm of the error operator appears extremely well correlated with the maximum nuclear charge in a molecule when using a local basis of atomic orbitals. The black line is the line of best fit for a $Z_\textrm{max}^6$ scaling.}
\label{fig:z_max_vs_error_norm}
\end{figure}

\fig{orbitals_vs_norm} indicates that Trotter error norm correlates especially well with the maximum nuclear charge, as further demonstrated in \fig{z_max_vs_error_norm}. The local basis is formed from the set of orthogonal atomic orbitals which are obtained for molecules using L{\"o}wdin symmetric orthogonalization on the original non-orthogonal local Gaussian orbitals \cite{McClean2014}. These Gaussian basis functions are constructed as approximations to eigenfunctions of Hydrogen-like systems, with some fitting adjustments. As such, we can determine the scaling behavior by considering the eigenfunctions of Hydrogen-like systems which are simple enough to permit analytical determination of how each term in the Hamiltonian will scale with nuclear charge. We begin by writing the eigenfunctions of a single electron in the potential of a point charge $Z$ in a convenient way,
\begin{align}
&\psi_{n \ell m} \left(\rho, \theta, \phi\right) = \\
&\sqrt{\left(\frac{2 Z}{n}\right)^3 \frac{\left(n - \ell - 1\right)!}{2 n \left(n + \ell\right)!}} e^{-\frac{\rho}{n}} \!\left(\frac{2 \rho}{n}\right)^{\ell}\! L^{2 \ell + 1}_{n - \ell - 1}\! \left(\frac{2 \rho}{n}\right)Y_{\ell}^m\!\left(\theta, \phi\right)\nonumber
\end{align}
where $\rho = r Z$, $L^{2 \ell + 1}_{n - \ell - 1} \left(\frac{2 \rho}{n}\right)$ is a generalized Laguerre polynomial of degree $n - \ell - 1$, and $Y_{\ell}^m\left(\theta, \phi\right)$ is a spherical harmonic of degree $\ell$ and order $m$. With the convention,
\begin{align}
\varphi_p \left(\sigma_i\right) & = \psi_p \left(\rho_i, \theta_i, \phi_i\right) \chi\left(s_i\right) \propto Z^{3/2}\\
d \sigma_i & =  \frac{\rho_i^2 \, d \rho_i}{Z^3} \sin\left(\theta_i\right)\,d\theta_i \,d \phi_i \,ds_i \propto Z^{-3}\\
\nabla^2 & = Z^2\left(\frac{\partial^2}{\partial \rho^2} + \frac{2}{\rho}\frac{\partial}{\partial \rho}\right) + \frac{Z^2}{\rho^2 \sin^2 \left(\theta\right)}\frac{\partial^2}{\partial \phi^2}\nonumber\\
& + \frac{Z^2}{\rho^2 \sin^2 \left(\theta\right)}\frac{\partial^2}{\partial \phi^2} \propto Z^2
\end{align}
where $\chi \left(s_i\right)$ is a spin assignment and $\sigma$ represents all degrees of freedom for an electron, we rewrite \eq{single_int} and \eq{double_int} in terms of $\rho$, assuming a single nuclei,
\begin{align}
h_{pq} & = \int d\sigma\ \varphi_p^*(\sigma) \left(-\frac{\nabla^2}{2} - \frac{Z^2}{\rho} \right)\varphi_q(\sigma)\\
h_{pqrs} & = \int  d\sigma_1 \, d \sigma_2 \frac{\varphi_p\left(\sigma_1\right) \varphi_q\left(\sigma_2\right)\varphi_s\left(\sigma_1\right) \varphi_r\left(\sigma_2\right)}{|\rho_1 - \rho_2| / Z}.
\end{align}
For both integrals, factors of $Z$ from the differential volume elements $d \sigma$ cancel with factors of $Z$ from the spin orbitals $\varphi$ and we find that,
\begin{align}
|h_{pq}| & = \Theta\left(Z^2\right)\\
|h_{pqrs}| & = \Theta\left(Z\right).
\end{align}

Thus, it is clear that we can upper bound the scaling of individual Hamiltonian terms with nuclear charge as ${\cal O}\left(Z_\textrm{max}^2\right)$. While this result is rigorous only when the orbital basis is the basis of true atomic orbitals, we expect qualitatively similar behavior in other bases. Assuming the $h_{pq}$ terms dominates the error in the Trotter formula then~\eq{error_op} implies that the Trotter error should scale as $\mathcal{O}(Z_{\rm max}^6)$.  This scaling is qualitatively consistent with the empirical scaling in~\fig{z_max_vs_error_norm} which fits $Z_\textrm{max}^6$ scaling to the norm of the error operator with an $r^2$-value of 0.994.  Comparable results to this scaling have also been observed in diffusion Monte Carlo algorithms~\cite{Ceperley1986,Hammond1987}.

\begin{figure}
\centering
\includegraphics[width=8cm]{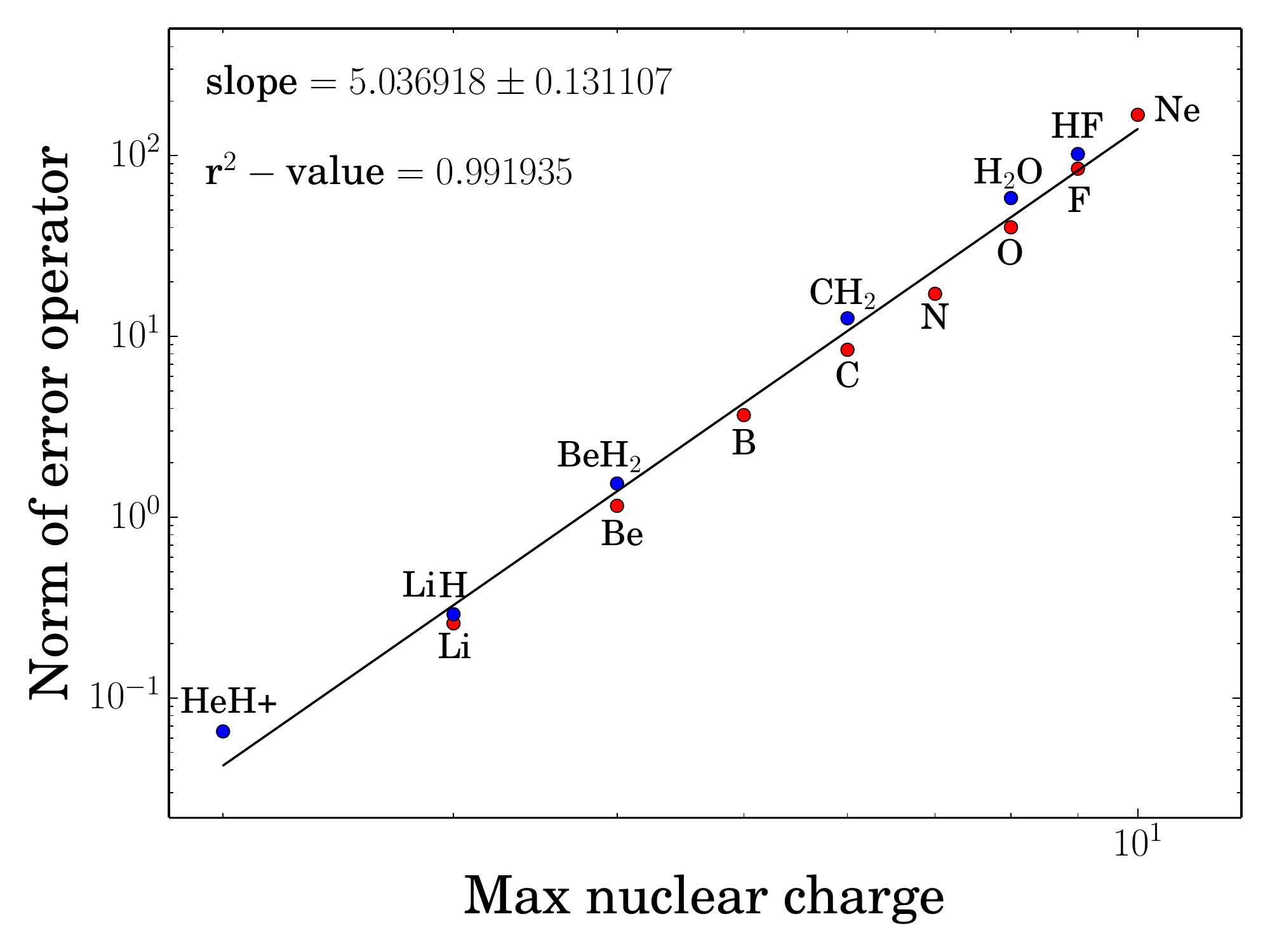}
\caption{The norm of the error operator appears also well correlated with the maximum nuclear charge in a molecule when using the canonical basis of molecular orbitals. The black line is a least squares fit to the data which is roughly consistent with a $Z_\textrm{max}^5$ scaling.}
\label{fig:z_max_vs_error_norm_CMO}
\end{figure}

These results imply that if an atomic basis is used then the error in the second-order Trotter-Suzuki formula scales at most as
\begin{equation}
\|V^{(1)}\|\in \mathcal{O}\left(N^4Z_{\rm max}^6+N^{10}Z_{\rm max}^3 \right).
\end{equation}
This result is a direct consequence of bounds on the Trotter-Suzuki error in~\cite{Hastings2014} and the observation that double commutators of the one- and two-body terms produce at most $N^4$ and $N^{10}$ terms respectively.
This implies that the computational complexity of performing the simulation on an arbitrary state, given fixed error tolerance of chemical accuracy, is $\mathcal{O}(N^4(N^2Z_{\rm max}^3+N^5Z_{\rm max}^{3/2}))$.  However,  our numerical results are consistent with an $\mathcal{O}(N^4Z_{\rm max}^3)$ which suggests that this scaling may be loose.  It also important to note that the gate depth can be further reduced by using interleaving and nesting as per~\cite{Hastings2014}, which is significant when the algorithm is implemented on systems where quantum operations can be executed in parallel.  It is also worth noting that the one-body terms dominate the two-body terms in every numerical example that we considered.  Larger molecules with more $h_{pqrs}$ terms may lead to Trotter errors that scale as $\mathcal{O}(Z_{\max}^3)$ rather than $\mathcal{O}(Z_{\max}^6)$.  More extensive numerical results may be needed to determine the conditions under which the two-body terms asymptotically dominate the one-body terms (if such conditions exist).

\fig{z_max_vs_error_norm_CMO} shows that these error estimates are pessimistic for the molecules considered when using the canonical basis. While the error norm is still strongly correlated to nuclear charge, unlike the scaling in the local basis, the fit to a $Z_{\max}^6$ scaling is less convincing. Instead, the data empirically seems to follow a $Z_{\max}^5$ scaling.  Intuitively, this is easy to envision because the molecular orbitals are inherently delocalized and thus it is natural to expect that the maximum nuclear charge should make less of an impact in this basis.  We also see no evidence of explicit scaling with $N$ over this range in $Z_{\max}$. It is interesting to note that although the number of non-negligible integrals in a local orbital basis can be quadratically or quartically smaller than the size of an untruncated canonical molecular orbital basis, the scaling with $Z_\textrm{max}$ seems to be better by a linear factor.  This suggests interesting trade-offs between the two methods and hints that neither is intrinsically superior for quantum simulation.

\subsection{Dependence on orbital structure}

\begin{figure}
\centering
\includegraphics[width=8cm]{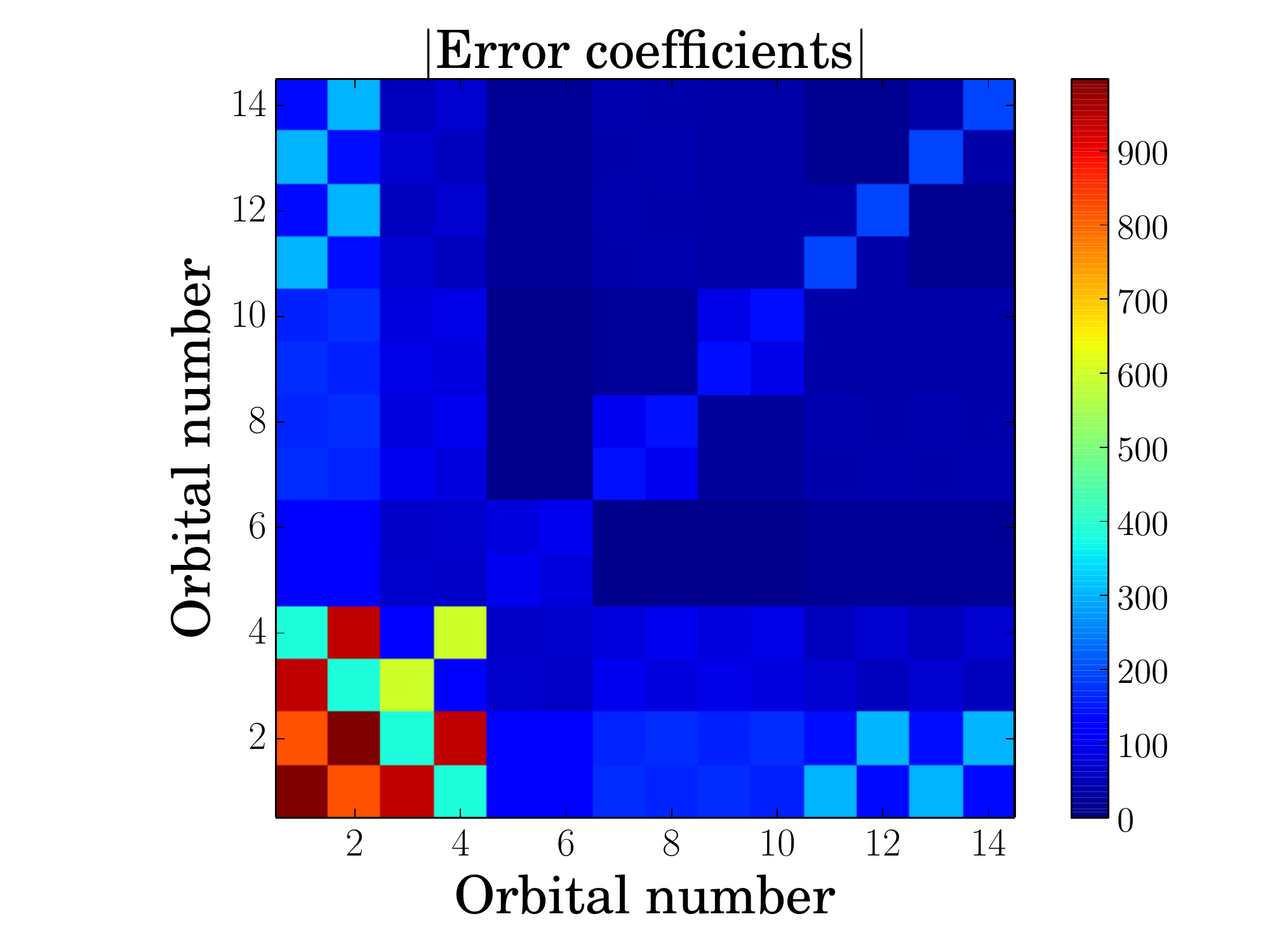}
\caption{This plot shows the coefficients of normal-ordered terms in the error operator for water in a local basis as a function of the orbitals on which they act. The coefficients of the error terms are binned according to the orbitals involved in the term. This plot shows the marginal distribution of the magnitudes of those terms.}
\label{fig:h2o_coeff_2d}
\vspace{-.5cm}
\end{figure}

The terms that appear in the error operator include interactions between every orbital in the basis set.  This begs the question of whether terms in the Hamiltonian that involve particular orbitals have larger contribution to the error.  In order to assess this, we compute the error operator for a number of different molecules and normal-order the resultant operator.  We then sum the magnitudes of every remaining term that either create or annihilate an electron in each of the orbitals. An example of this is provided in \fig{h2o_coeff_2d}, which shows the marginal coefficient magnitudes of all terms in the error operator (after normal-ordering) in terms of two spin orbitals they contain. \app{extradata} shows similar analysis for other molecules in other basis sets. As we can see, terms which involve the inner shell electrons dominate the norm of the error operator in the local basis.

We see from such figures that the inner orbitals, especially the single particle terms which are on the diagonal of the plot above, have a substantial impact on the Trotter error.  This is not surprising as the inner atomic orbitals interact very strongly with nuclei so the single particle integrals are likely to be much larger than the interaction integrals for these orbitals.  Interestingly, although the valence shell electrons are often the most important for determining the chemical properties of a molecule, the inner orbitals are the ones that affect the error most significantly.  This suggests that pseudo-potentials, which allow the core electrons to be treated as effectively ``frozen'', may  provide a way to reduce the Trotter error in some circumstances. We leave this as an open question for future work.

\subsection{Dependence on structure of eigenstates}

Due to the substantial discrepancy between error induced on the ground state and operator norm, we might ask the following question: given the error operators for real molecules, what is the distribution of errors that would be induced on a random ensemble of vectors? This question is important as the answer will help us to identify the source of the observed error cancellation. We consider the ensemble of Haar random vectors which form a unitarily invariant ensemble of vectors with uniformly distributed complex elements. Unitary invariance ensures that the ensemble has uniform distribution in an arbitrary complete, orthonormal basis such as the eigenbasis of the error operator.

\begin{figure}
\centering
\includegraphics[width=8cm]{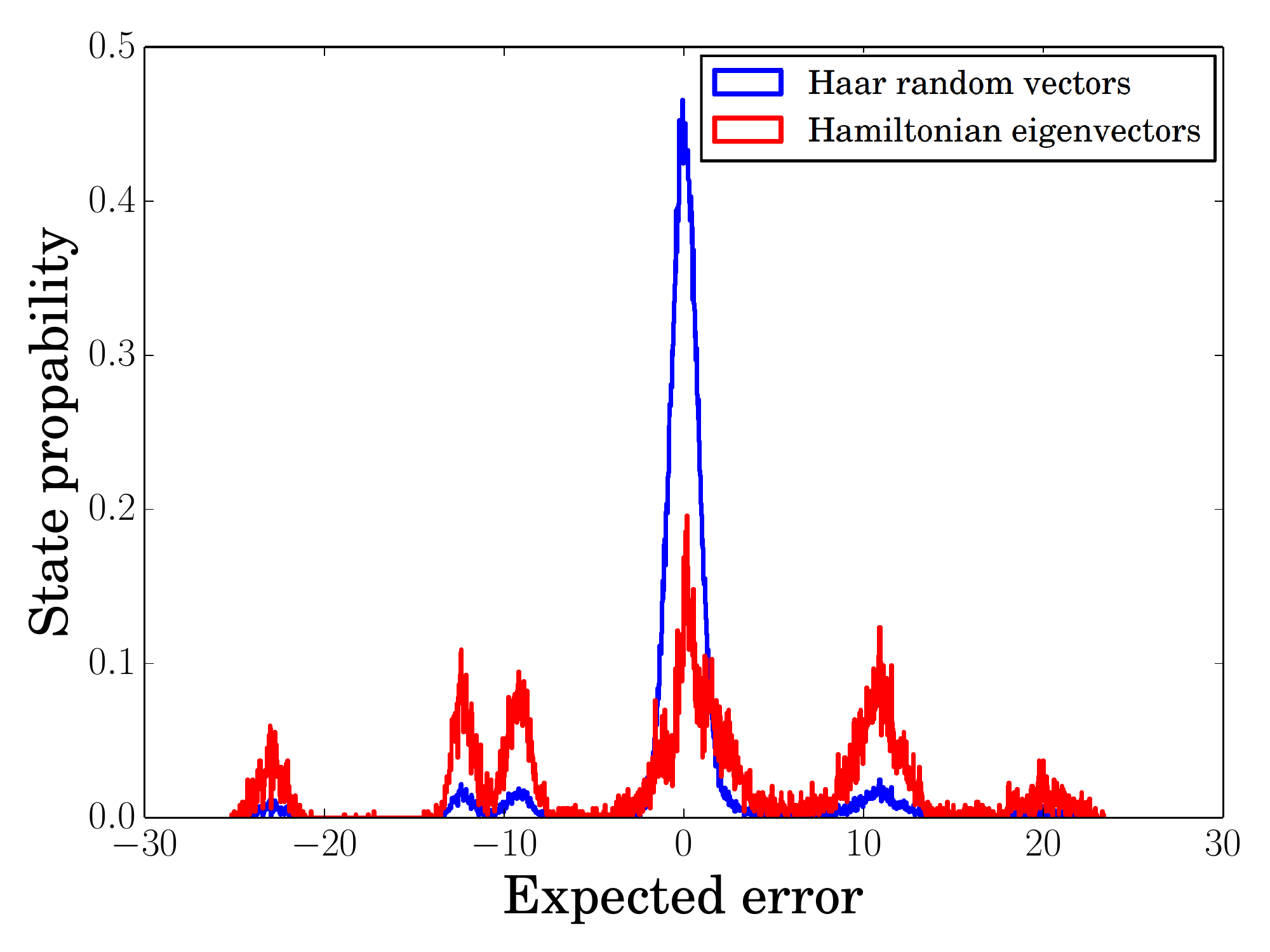}
\caption{This plot shows the distribution of expectation values of the error operator for water in the local basis over its eigenstates and Haar random vectors. We see that the random vectors lead to substantially less error, on average, than do the Hamiltonian eigenstates. The Haar distribution of errors has a standard deviation of 4.82 while the Hamiltonian error distribution has standard deviation of 10.68.}
\label{fig:H2O_OAO_expectations}
\end{figure}

\begin{figure*}
         \centering
         \subfloat{
                 \includegraphics[width=.3\textwidth]{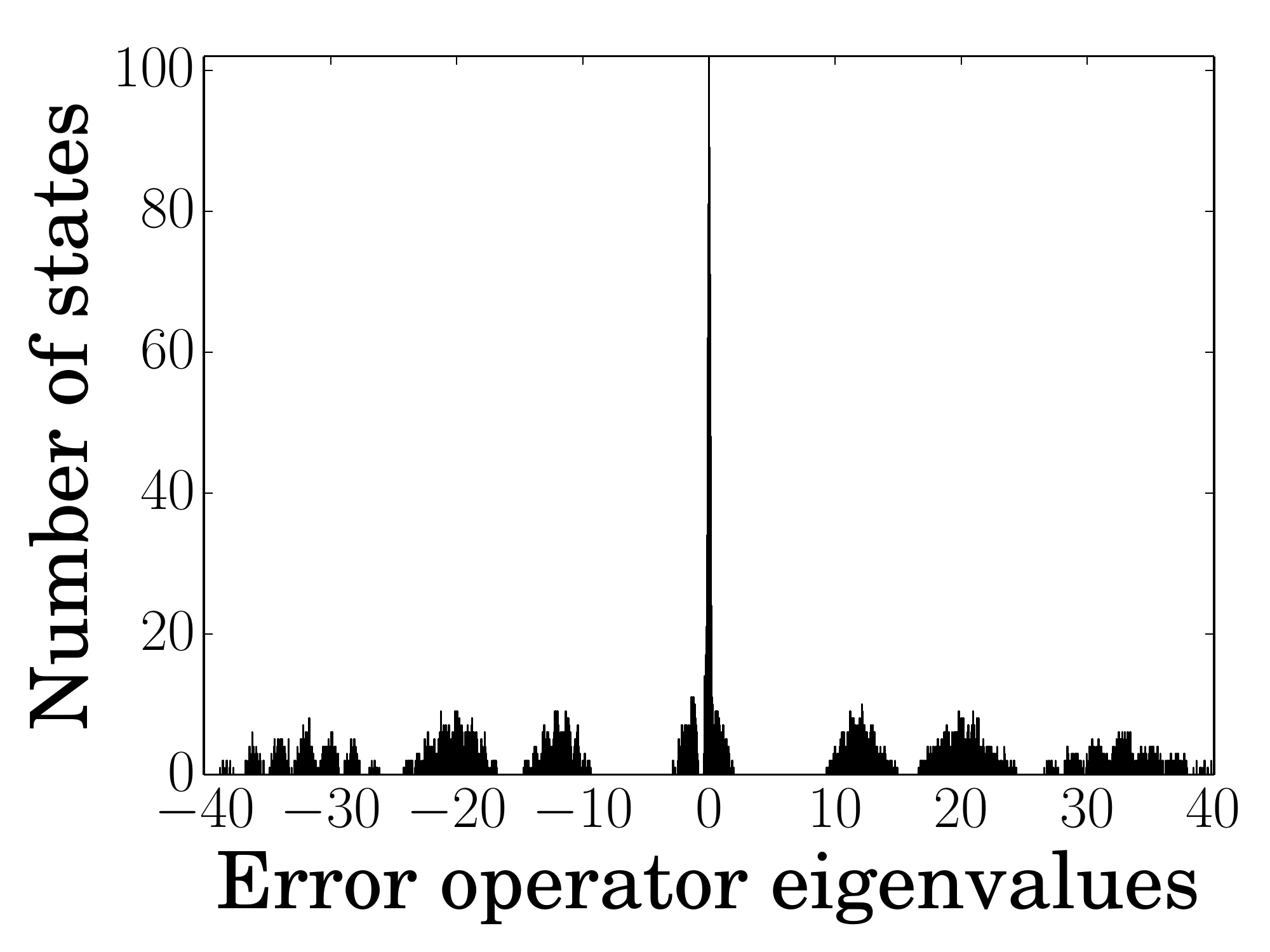}}
         \subfloat{
                 \includegraphics[width=.3\textwidth]{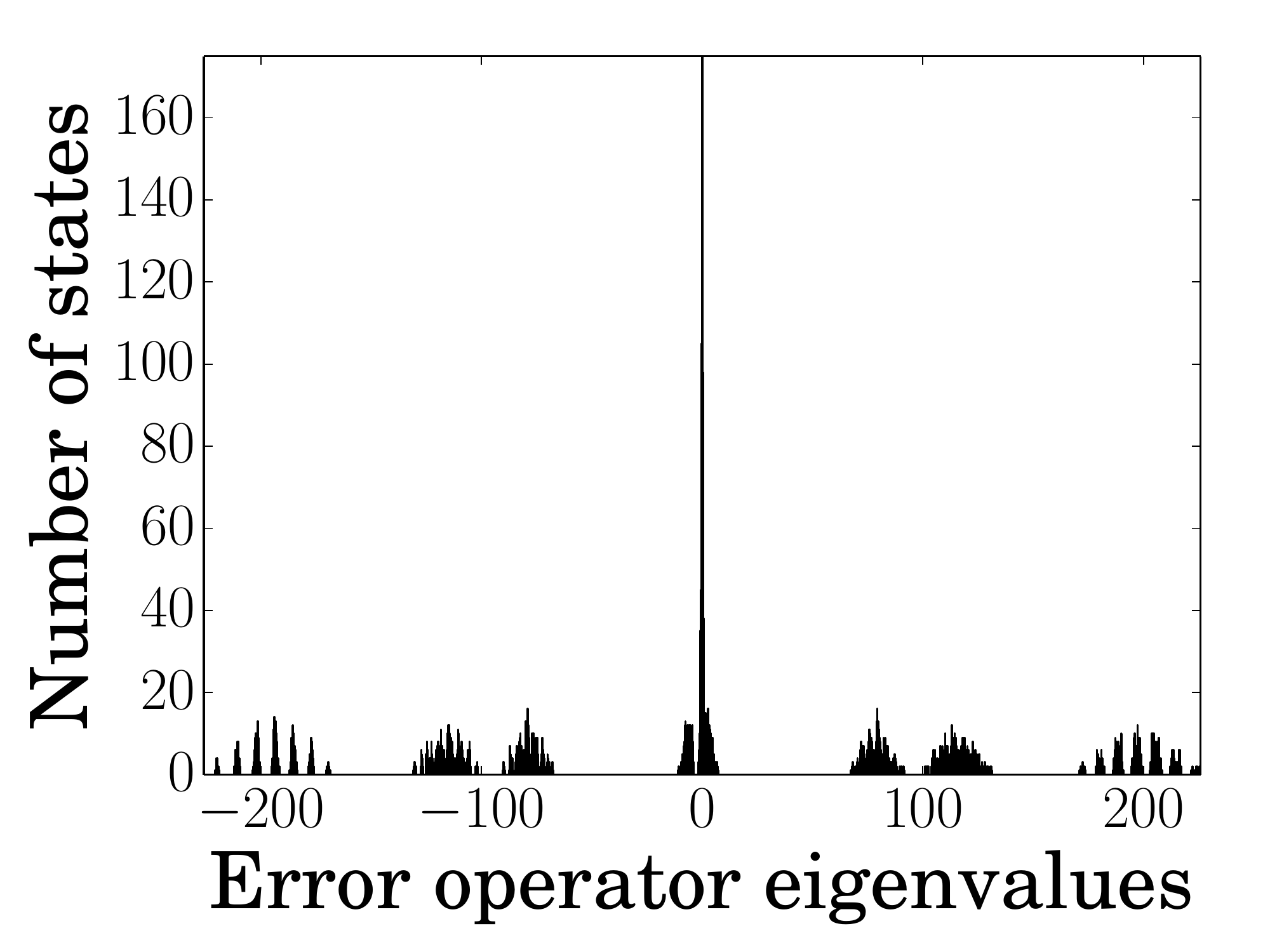}}
         \subfloat{
                 \includegraphics[width=.3\textwidth]{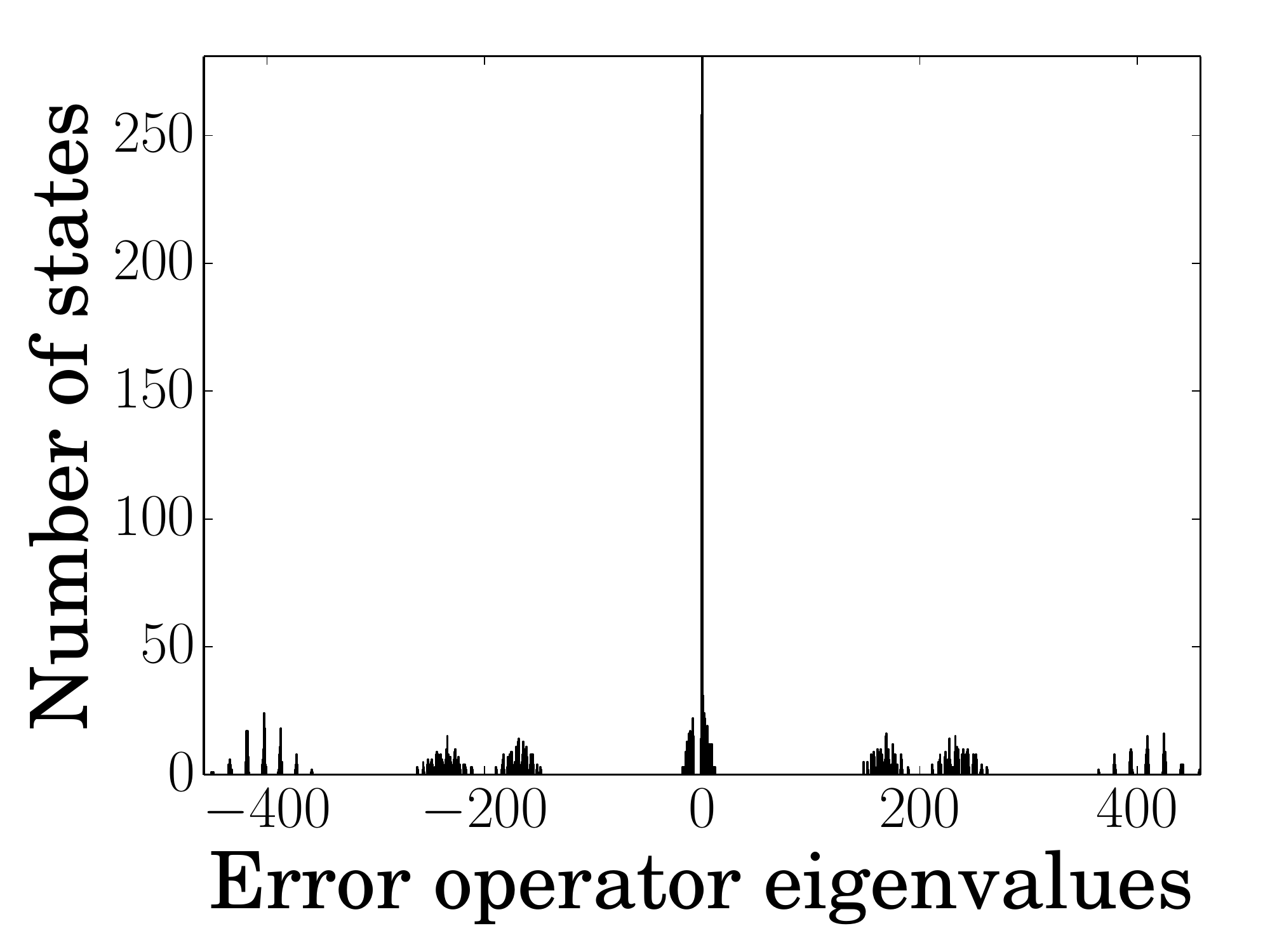}}\\
         \subfloat{
                 \includegraphics[width=.3\textwidth]{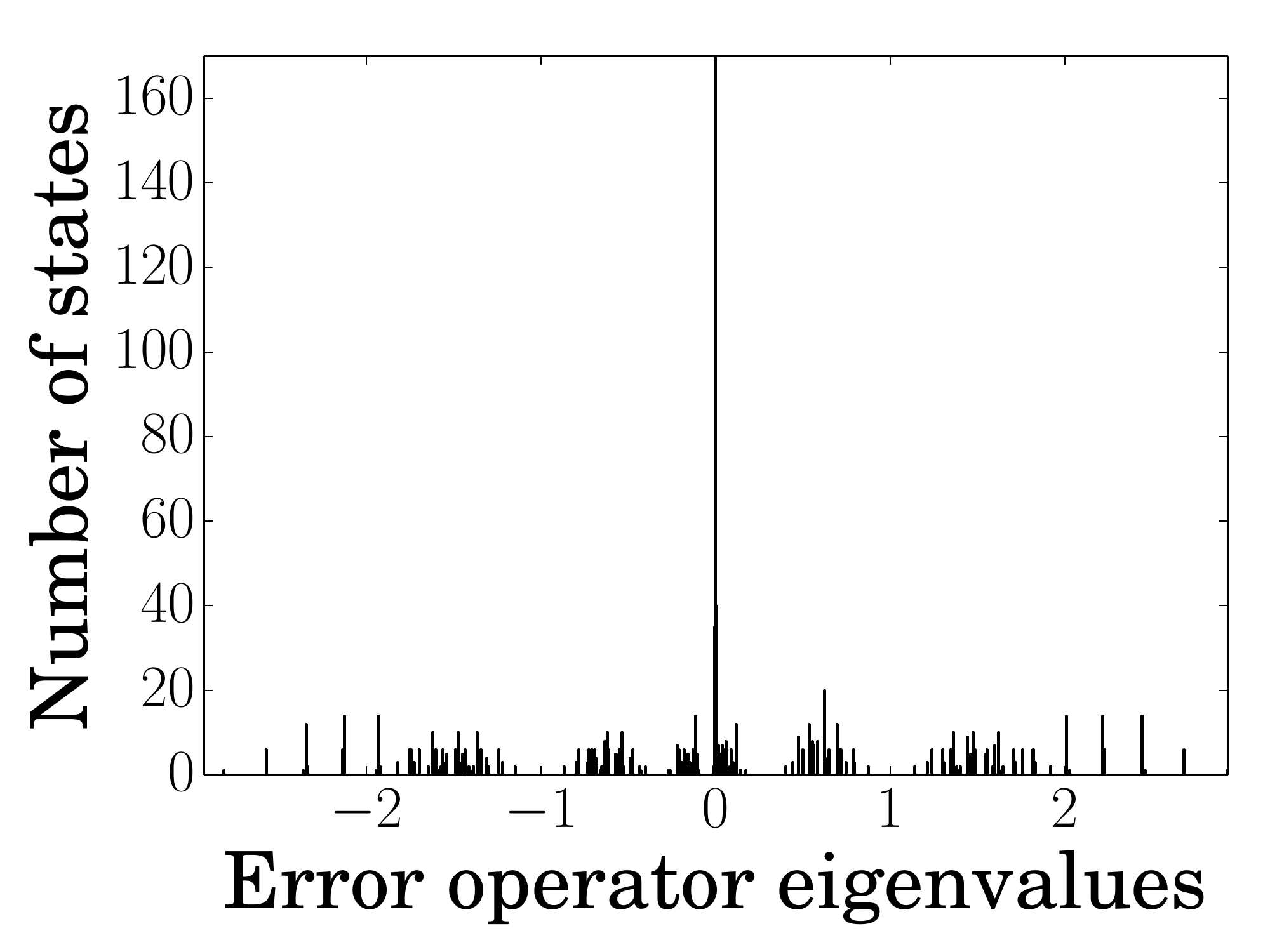}}
         \subfloat{
                 \includegraphics[width=.3\textwidth]{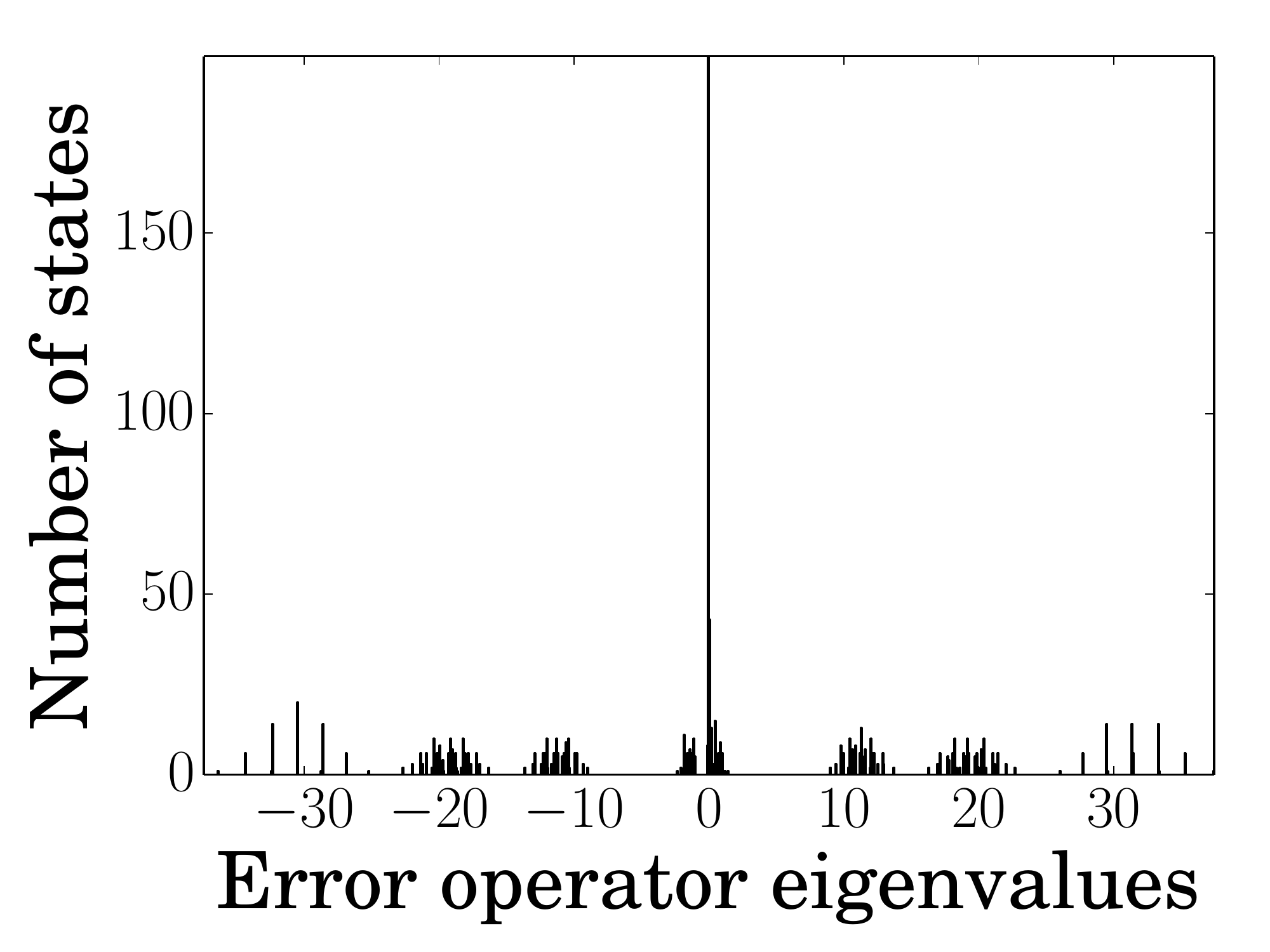}}
         \subfloat{
                 \includegraphics[width=.3\textwidth]{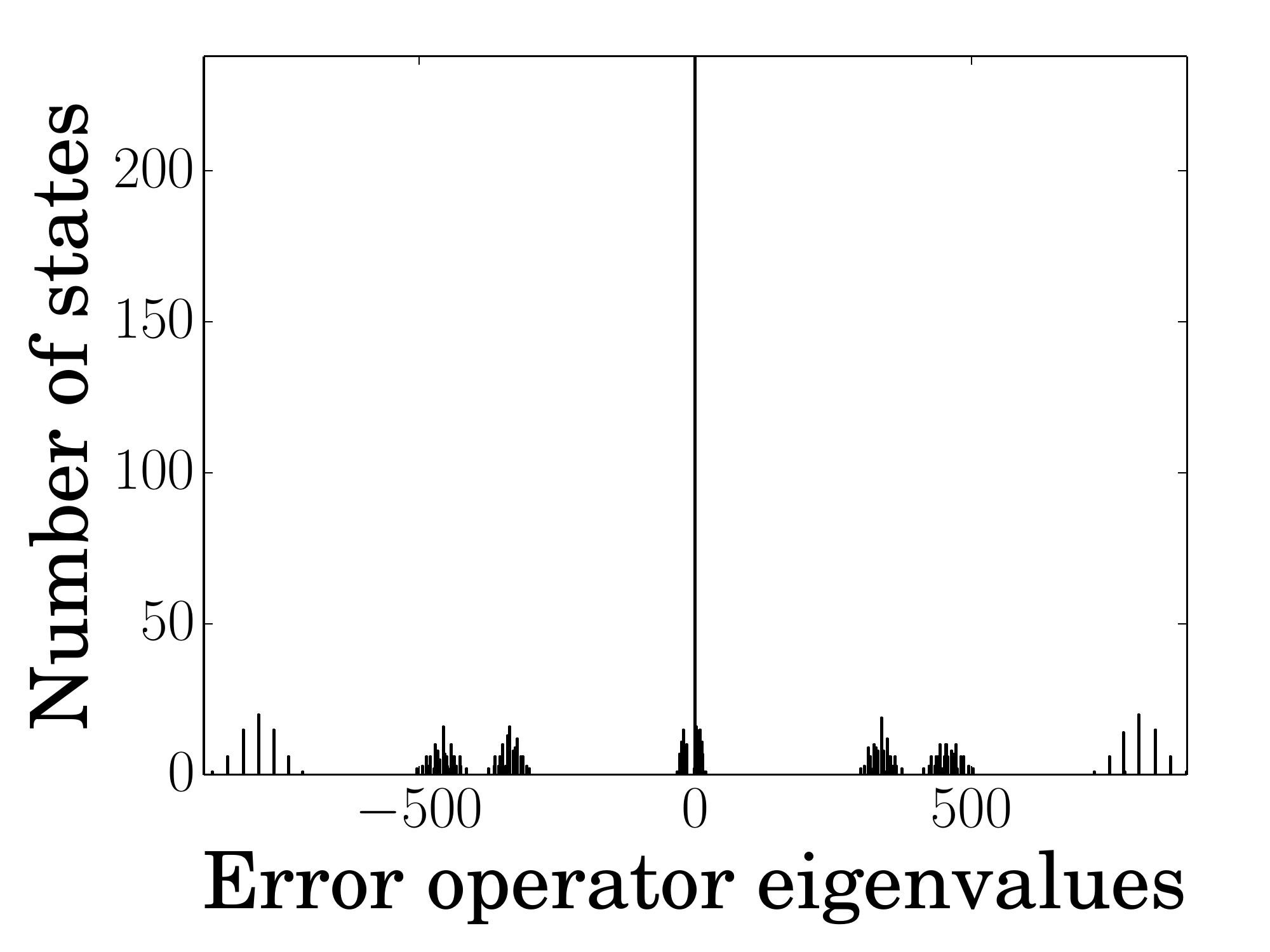}}
         \caption{These are histograms of the eigenspecta of the error operators for various molecular and atomic benchmarks in the local basis. Proceeding clockwise from the top left, the molecules are water, Hydrogen Fluoride, Methylene, atomic Beryllium, atomic Carbon and atomic Neon. Error operators for all of our benchmarks have surprisingly similiar eigenspectra, regardless of the orbital basis. The source of this striking similarity and the reason for the particular structure is unknown.}\label{fig:errorgenspectra}
\end{figure*}

Denoting vectors from the random ensemble as $\ket{v}$ and eigenvectors of the error operator as $\ket{k}$ with eigenvalue $\lambda_k$, we are interested in analyzing properties of the following distribution of expected errors given by,
\begin{align}
\Delta E\left(v\right) = \sum_k \lambda_k |\langle v \ket{k}\!|^2.\label{eq:ex_errors}
\end{align}
First, note that $\sum_k \lambda_k =0$.  This is because if $C = [A,B]= \sum_j \lambda_k \ket{k}\!\!\bra{k}$ then 
\begin{equation}
\sum_k \lambda_k = \rm{Tr}(C)=\rm{Tr}(AB) -\rm{Tr}(BA)=0,
\end{equation}
from the cyclic property of the trace. Since $V^{(1)}$ is the sum of such operators, it follows that its trace is also zero.  This implies that the Haar-expectation value of the error, over all possible random states, is
\begin{equation}
\mathbb{E}_H (\Delta E(v))=\sum_k \lambda_k \mathbb{E}_H |\langle v \ket{k}\!|^2 = \frac{1}{2^N}\sum_k \lambda_k=0.
\end{equation}
This shows that there is no inherent bias that arises from Trotterization towards either overestimating or underestimating the true expectation value.

This result does not represent the \emph{typical error} that we expect to see in a simulation.  We also need to find the Haar variance of the expected error to estimate the typical variation of simulation errors about the mean.  It is then easy to see that the Haar variance is
\begin{equation}
\mathbb{V}_H(\sum_k \lambda_k |\braket{v}{k}\!|^2) = \sum_k \lambda_k^2 \mathbb{V}_H(|\braket{v}{k}|^2).\label{eq:VH}
\end{equation}
In \app{Haar}, we  derive the Haar  variance of the squared projection,
\begin{align}
\mathbb{V}_H\left(|\langle v \ket{k}|^2\right) =  \frac{2}{2^N \left(2^N + 1\right)} - \frac{1}{2^{2 N}},
\label{eq:sigma_p}
\end{align}
where $N$ is the number of spin orbitals. Combining \eq{sigma_p} and~\eq{VH} and using Chebyshev's inequality, we see that with high probability over $\ket{v}$
\begin{equation}
|\bra{v} V^{(1)}\ket{v}| \in O\left( \frac{\sqrt{\sum_k \lambda_k^2}}{2^N} \right).\label{eq:concentration}
\end{equation}
\eq{concentration}, surprisingly, shows that a concentration of measure argument causes the expectation of the Trotter error to be asymptotically zero if ($a$) $\ket{v}$ is typical of a Haar random state, (b) $\sum_k \lambda_k^2 \in o(2^{2N})$ and ($c$) $\ket{v}$ is chosen independently of the $\ket{k}$.

We do not expect a concentration of measure argument like this to hold for actual quantum simulations because it would imply that the Trotter errors \emph{in eigenvalue estimation} shrink rapidly with system size for physically reasonable distributions of $\lambda_k$.  Thus, it is natural to expect that one or both of assumptions ($a$) and ($c$) are not reasonable for eigenvalue estimation.

In \fig{H2O_OAO_expectations}, we show the expected errors according to \eq{ex_errors} over an ensemble of Haar random vectors as well as the expected errors over the eigenstates of the Hamiltonian for water. The results clearly show that the errors observed in this chemical example are much greater than we would expect from Haar random states.  Furthermore, we see little evidence of concentration of measure of the errors about zero for the case where $\ket{v}$ is an eigenvector of $H$; whereas the Haar random $\ket{v}$ lead to results that are much more concentrated about zero error.  This suggests that the discrepancies between the norm of the error operator and the ground state error cannot be explained by a simple randomization argument as the actual errors observed are much worse than would be otherwise expected.

\eq{ex_errors} shows that the expected error is the convolution of the functions $\lambda_k$ and $|\braket{v}{k}|^2$.
Thus, we expect the distribution of errors to resemble the underlying distribution of eigenvalues of $V^{(1)}$.  This intuition can easily be seen by comparing~\fig{H2O_OAO_expectations} to the eigenspectrum of the water error operator in  \figa{errorgenspectra}{a}. As expected, the distribution of errors for the random ensemble (\fig{H2O_OAO_expectations}) resembles the error operator eigenspectrum (\figa{errorgenspectra}{a}) with concentration about the mean (as anticipated by~\eq{concentration}).  Also, it is interesting to note that the eigenspectra of the error operators for various molecules and atoms studied in this paper bear a remarkable degree of similarity and appear extremely structured as \fig{errorgenspectra} demonstrates.
Additionally, every example has a sharp peak in its spectrum about zero error.  This suggests that much of the rift between the norm of the error operator in~\fig{basis_set_compare} may be due to the  large number of eigenvectors with near-zero eigenvalue.

\section{Improved simulation methods inspired by classical approaches to quantum chemistry}

Given the large disparity between error operator norm and error induced on the exact ground state, any efficient method which allows one to approximate the error induced on the ground state (which implies an estimate for the number of Trotter steps needed) would be of critical importance for anyone wishing to actually run a quantum chemistry simulation on a quantum computer.  A natural way to address this problem is to directly evaluate the error over a mesh in position and fit the data to a power law.  This process can be made efficient using the SWAP test, as proposed by Wiebe et al~\cite{WBH+11}.  A major drawback of this approach is that it requires roughly twice the qubits that the basic simulation used and also the variance in the estimate returned by the SWAP test can be prohibitively large.
In this section, we propose an alternative method that estimates the error in the ground state energy by evaluating the error operator on a classical ansatz for the ground state numerically.  This method also allows the contribution to the error in the quantum simulation from the Trotter error to be subtracted off of the final estimate, improving the accuracy of the simulation without requiring additional quantum operations.

Perhaps the most well-known classical algorithm for solving the electronic structure problem is a mean-field approach known as the Hartree-Fock method  \cite{Helgaker2013}. In this scheme, single particle molecular orbitals are obtained using a self-consistent variational procedure in which each particle is made to interact with the average density of the other particles. The output of this calculation provides molecular orbitals which, together with a spin assignment, are used to approximate the $n$-particle wavefunction as an anti-symmetric product of the orbitals (known by chemists as a Slater determinant).

Unfortunately, the Hartree-Fock method is incapable of approximating dynamic electron correlation and is known to overestimate energies by an amount that is typically well above the threshold of chemical accuracy. To correct for this problem, one can expand the wavefunction in a basis of multiple Slater determinants and variationally solve for the coefficients which minimize the electronic energy. In general, there are $M = \binom{N}{n}$ valid configurations for $n$ electrons arranged into $N$ spin orbitals. The ground state wavefunction in \eq{electronic} may be represented as a linear combinations of these arrangements,
\begin{align}
\label{eq:fci_expansion}
\ket{\Psi} = \sum_{i=1}^{M} a_i\ket{i}.
\end{align}
The energies may be solved for variationally,
\begin{align}
E = \min_{\left\{a_i\right\}} \frac{\bra{\Psi} H \ket{\Psi}}{\braket{\Psi}{\Psi}} \quad \rightarrow \quad H^\textrm{CI}\ket{\Psi} = E \ket{\Psi}
\end{align}
where $H_{ij}^\textrm{CI} = \bra{i} H \ket{j}$. In chemistry this method is known as full configuration interaction (FCI).

FCI is strongly believed to be classically intractable because $M$ scales combinatorially with $N$ and $n$. Accordingly, a common classical approach is to truncate the expansion in \eq{fci_expansion} to include only configurations that represent a fixed number of excitations from a reference configuration. Though this work and recent work \cite{McClean2014} discuss using different orbital basis choices, usually the reference is taken to be the Hartree-Fock state (this orbital basis is known in chemistry literature as the ``canonical basis''). This approach defines a hierarchy of methods referred to as truncated configuration interaction (CI) which approach exactness as the number of excitations is increased to the FCI space spanned by $N - n$ excitations. Fixing the maximum number of excitations at $k$, combinatorics suggests that the number of basis functions in truncated CI scales as $\Theta\left({N - n \choose k}{n \choose k}\right)$. Truncation to the level of single and double excitations is referred to as configuration interaction singles, doubles (CISD) and is used for several purposes in this paper. Finally, we note that the accuracy of truncated CI is extremely sensitive to the quality of the reference state and it is therefore difficult to determine when these methods are expected to approximate the ground state energy within even a fixed multiplicative error.

\begin{figure}
\centering
\includegraphics[width=8cm]{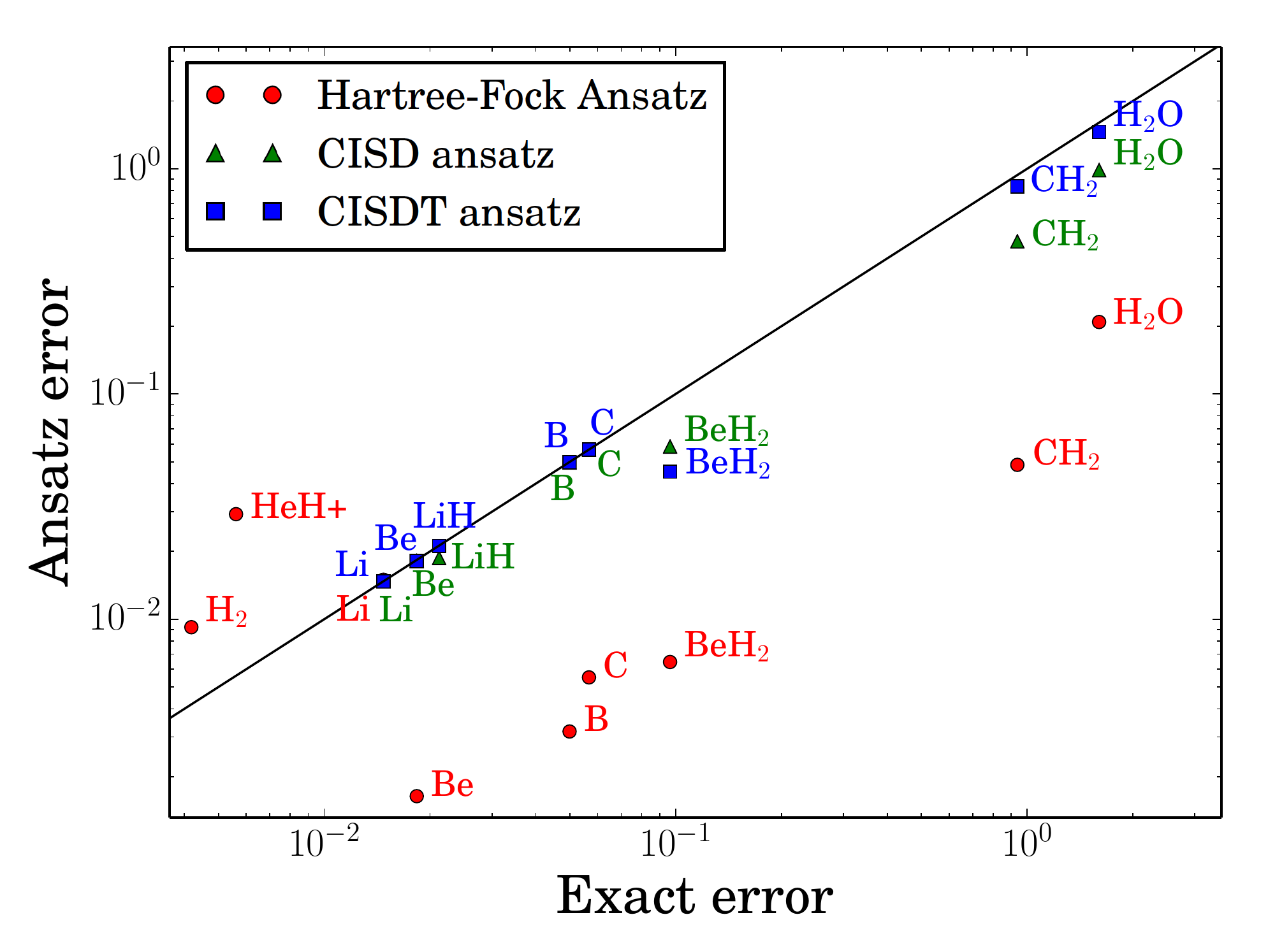}
\caption{Magnitude of the Trotter error in the exact ground state against the magnitude of the error induced on a classical ansatz for the ground state. Truncated CI computations are only performed when inexact; e.g., we have not computed HF using CISD because the calculation is exact in STO-6G.}
\label{fig:configuration_compare}
\end{figure}

\begin{figure}
\centering
\includegraphics[width=8cm]{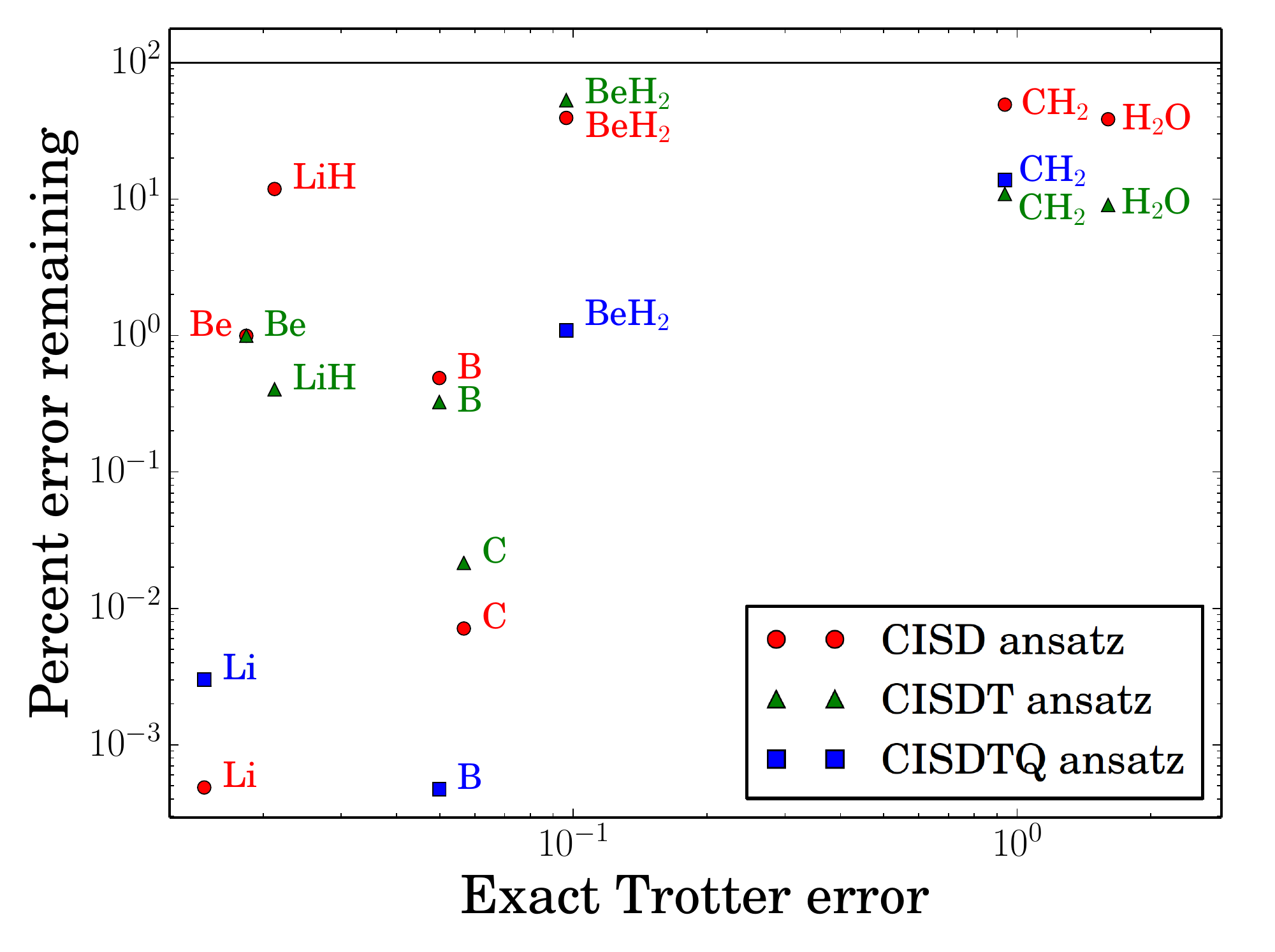}
\caption{Magnitude of the Trotter error induced in the exact ground state against percentage of the error that remains after subtracting the ansatz error from the exact error. This plot is intended to indicate the reduction in effective error when using a classical ansatz estimate. A black line is drawn at one-hundred percent remaining. In all benchmarks, using these ansatzs reduces effective error. Note that the quadruple calculation is so accurate for Be and LiH that the effective error appears to be exact to within double precision.}
\label{fig:configuration_reduction}
\end{figure}

Since the error operator can be efficiently computed and normal-ordered in second quantized form, we suggest evaluating the expectation value of this operator on a classical ansatz for the ground state. In particular, we focus on the use of the configuration interaction ansatz. \fig{configuration_compare} illustrates the utility of this idea by showing the discrepancy between actual error and the error from evaluation of the error operator using a classical ansatz. \fig{configuration_reduction} shows the extent to which the effective error is reduced using a classical ansatz.

Apart from estimating errors, CISD states may also be of use in coalescing schemes~\cite{Poulin2014} which use the Hartree-Fock approximation to determine whether a term in the Hamiltonian can be executed less frequently without significantly impacting the quality of the simulation.  This process can substantially reduce the costs of simulating molecules with many small, but non--negligible, $h_{pqrs}$ terms but may fail if  the Hartree-Fock approximation breaks down.  In such cases, the use of CISD states may lead to superior coalescing schemes at the price of requiring more classical computing time to find the coalescing schedule.

Though the Hartree-Fock ansatz is usually not accurate enough to reduce error by an order of magnitude, the use of a truncated CI ansatz often exhibits enough accuracy to very substantially reduce effective error. While we focus on the CI ansatz to provide proof-of-principle,  we believe that more intelligent truncation schemes can substantially increase ansatz accuracy without additional computational cost. For instance, the use of multi-reference methods has been shown to greatly improve the quality of the classical solution in many cases, especially near molecular dissociation limits where the exact electronic states become nearly degenerate \cite{Helgaker2013}.

The idea of using a classical ansatz to reduce the effective error in a quantum calculation is useful for two reasons. The first reason is that the error in a quantum simulation can usually be reduced by approximating the error with a classical ansatz at the CISD level of theory or greater, as demonstrated in \fig{configuration_reduction}.  The second (and perhaps more important) reason this technique is useful is that it gives a realistic a priori estimate of the error to expect in the quantum simulation (expected to be correct to at least an order of magnitude) which provides a methodology for selecting the number of Trotter steps required to obtain a desired precision. Finally, we point out that while the error operator might be computationally costly to compute  (albeit, efficient in the polynomial-time scaling sense), Monte Carlo methods could be used to tractably sample the error operator expectation values with a classical ansatz.

\subsection{Circuit for state preparation based on CI ansatz}

In contrast to the Hartree-Fock states, CISD states are not computational basis states.  Instead they are a linear combination of quantum states that are formed by single and double excitations away from a reference state which is often taken to be the Hartree-Fock state.  Although the CISD state can be efficiently computed for a given electronic structure problem, preparing the state on a quantum computer is non-trivial.  Here we present a method based on state-of-the-art multi-qubit synthesis methods to prepare the CISD state.  Previous work has considered preparing this state using single qubit rotations and CNOT gates~\cite{OGK01,WAF09,Yung2013}.  Such gate sets are unrealistic for fault tolerant quantum computing so we discuss the problem of compiling the state preparation circuit into Clifford and $T$ gates.  In the following analysis we will take the cost of the circuit to be given by the number of $T$ gates because these gates are  the most expensive gates to implement fault tolerantly in error correcting codes such as the surface code.

Let us begin by assuming the initial state for the quantum simulation (i.e. the state we wish to prepare) is of the form
\begin{equation}
\ket{\psi} =\sum_{k=1}^D \alpha_k \ket{j_k},
\end{equation}
where $j_k$ is a sequence of computational basis vectors that spans the space that state has support over and $D$ is the dimension of that space.

It is unrealistic to assume that the state $\ket{\psi}$ will be exactly preparable using gates from the Clifford + $T$ gate library.  Instead, the initial state will typically have to be approximated using these circuit elements.  For years the Solovay-Kitaev algorithm provided the best known method for solving this approximation problem, but recently more advanced methods based on number theoretic results have provided much more efficient ways of performing this decomposition~\cite{KMM13,RS14,BRS14}.

Therefore the problem of finding the best sequence of Clifford and $T$ gates to approximate a multi-qubit unitary reduces to the following problem

\begin{enumerate}
\item Find integers $x_0,x_1,y_0,y_1$ such that $$U_{p,q}\approx \tilde U_{p,q}= \frac{x_0 +x_1 \sqrt{2}+ iy_0+iy_1\sqrt{2}}{\sqrt{2}^m}.$$ and $\tilde U$ is a unitary that can be exactly synthesized using elements from the gate library.
\item Find a sequence of Clifford and $T$ gates that exactly implements $\tilde{U}$.
\end{enumerate}
Note that because we are interested in preparing a state, not implementing a multi-qubit unitary, only the first column of $U$ needs to be approximated.  In particular, the first column
of $\tilde U$ should approximate $\ket{\psi}$ to within a fixed error tolerance $\delta$.

Before proceeding it is necessary to briefly review number theoretic approaches to multi-qubit circuit synthesis using Clifford and $T$ gates.  The key insight behind this strategy is that the unitary matrices that can be prepared with such circuits take on a very special form.  The form can easily be seen from the Hadamard and $T$ gates,
\begin{equation}
H = \frac{1}{\sqrt{2}} \begin{bmatrix} 1&1\\1&-1 \end{bmatrix}, \quad \quad T = \begin{bmatrix} 1&0\\0&\frac{1+i}{\sqrt{2}} \end{bmatrix}.
\end{equation}
It is then clear that any unitary matrix formed by a sequence of $H$ and $T$ gates will consist of matrix elements that are of the form
\begin{equation}
\tilde U_{i,j}=\frac{x_0 +x_1 \sqrt{2}+iy_0 +iy_1 \sqrt{2}}{\sqrt{2}^m},\label{eq:frac}
\end{equation}
for integer $x_0,x_1,y_0,y_1$.  Since the remainder of the gate set consists of CNOT gates and Pauli gates which have (complex) integer valued matrix elements, it is then clear that  every unitary that can be formed by the gate library also has matrix elements whose denominators are powers of $\sqrt{2}$ and whose numerators are in the ring of Gaussian integers $\mathbb{Z}[1/\sqrt{2},i]$.

Just like ordinary fractions, these fractions also can be reduced.  This notion of reducing a fraction manifests itself as the least denominator exponent $k$.  In order to understand this concept concretely, it is necessary to introduce some terminology.  Let $\omega=e^{i\pi/4}$ and 
\begin{equation}
\mathbb{Z}[\omega] = \{a\omega^3 + b\omega^2+c\omega+d| a,b,c,d \in \mathbb{Z}\}.
\end{equation}
Similarly, if we let $\mathbb{D}=\{a2^{-b}| a,b\in \mathbb{Z} \}$ denote the ring of dyadic fractions then we can express the ring $\mathbb{Z}[1/\sqrt{2},i]$  as
\begin{equation}
\mathbb{D}[\omega] = \{a\omega^3 +b\omega^2+c\omega+d| a,b,c,d \in \mathbb{D} \}.
\end{equation}
Then for every $t\in\mathbb{D}[\omega]$ there is a notion of a \emph{least denominator exponent} that describes the fraction in~\eq{frac} and uses the smallest value of $m$ possible while requiring that $x_0,x_1,y_0,y_1$ are integer.  Or more formally, the least denominator exponent, $k$, is the smallest non-negative integer such that $t\sqrt{2}^k \in \mathbb{Z}[\omega]$.

The smallest denominator exponent measures the precision in the approximation $U\approx \tilde{U}$ because~\eq{frac} allows arbitrary complex numbers to be represented with zero error in the limit as $k\rightarrow \infty$.  This means that the value of $k$ used in the rounding process of the first column of $U$ is a key property for characterizing the complexity of the state preparation.  In fact, the problem of bounding the error in this approximation problem as a function of $k$ has already been solved by Kliuchnikov~\cite{Kli13}:
\begin{equation}
\| (U-\tilde U)\ket{0}\| \le 2(D+2)2^{-4k} + 2\sqrt{2(D+2)}2^{-2k},\label{eq:normeq}
\end{equation}
where $D$ is the number of nonzero components of the state $\ket{\psi}=U\ket{0}$.  As a technical point, the dimension of $\tilde{U}$ is at most $D+2$ rather than $D$ because the first column of $U$ must have \emph{at least two zero-valued components} in order to guarantee that a solution exists to the Diophantine equation for $\tilde U$.  This requires enlarging the Hilbert space dimension by two in the worst case scenario, which may require adding at most an additional qubit.  However, the CISD state vector will likely have many zero valued components so this extra qubit will often not be needed in practice.

Using~\eq{normeq} we see that the state preparation error can be made less than $\delta$ by choosing
\begin{equation}
k=\left\lceil\frac{1}{4} \left[1+ \log_2\left(\frac{D+2}{(\sqrt{1+\delta}-1)^2} \right) \right] \right\rceil.\label{eq:keqn}
\end{equation}
This means that if $D$ is polynomial in $n$ then $k\in O(\log(n/\delta))$.

Once the unitary $\tilde U$ has been found then the task of decomposing the unitary into fundamental operations remains a non-trivial problem.  This problem is addressed by Giles and Selinger in~\cite{GS13}.  The idea behind this approach is to decompose $\tilde U$ into a series of two level unitary operations.  These two level unitary operations are then implemented using a Clifford circuit and a series of controlled operations to map each two level subspace to a single qubit.  This process involves first identifying pairs of levels that can be simplified and then performing circuits of the form $H^{w}T^xH^yT^z$ to the two level subspace such that the denominator exponent is systematically reduced.  Once the least denominator exponent is reduced to $0$ then the subspace either takes the form $[\omega^p,0]^T$ or $[0,\omega^p]^T$ for integer $p$.  Thus, the inverse of the state preparation circuit can be found (up to a global phase) by performing this reduction process iteratively of the $D$ dimensional initial state until only one nonzero component remains and then mapping this component to $\ket{0}$ using a Clifford circuit and a multiply controlled not gate.

At most $k$ reduction steps are needed to reduce each two level subspace and there are at most $(D+2)-1$ subspaces that must be looped through.  Therefore, there are at most $k(D+1)$ reduction steps taken.  Each reduction step consists of applying at most two $H$ gates and two $T^x$ gates to each subspace, as well as a multiply controlled not gate to map the final state to one proportional to $\ket{0}$.  Hence, in order to assess the cost of the algorithm we need to compute the costs of each of these gates.

Let us imagine that we need to perform a gate on the subspace ${\rm span}(\ket{j},\ket{k})$.  We want to map this to ${\rm span}(\ket{2^n-1},\ket{2^n-2})$ so that the gate can be applied to the last qubit.  By performing a sequence of $\mathcal{O}(n)$ $X$ gates, we can map $${\rm span}(\ket{j},\ket{k})\rightarrow{\rm span}(\ket{j\oplus k \oplus 2^n -1},\ket{2^n-1}),$$
where $\oplus$ is  bitwise exclusive or.  There are two cases that we need to consider.  If $j\oplus k = 1 \mod 2$ then the least significant bit of $j\oplus k \oplus 2^n -1$ is $0$.  This means that the state $\ket{j\oplus k \oplus 2^n -1}$ can be mapped to $\ket{2^n -2}$ using a sequence of $n-1$ zero-controlled not gates while not affecting $\ket{2^n-1}$.  Otherwise, if $j\oplus k = 1 \mod 2$ then we can reduce this case by finding the least significant bit where $j$ and $k$ differ and swap that bit with the least significant bit.  Since $\ket{2^n-1}$ is an eigenstate of the swap operator, the swap does not affect that vector.  Hence in either case we can perform the subspace mapping using $\mathcal{O}(n)$ Clifford operations.

In order to apply the $H$ and $T$ gates required by the synthesis algorithm on the correct qubits, we need to implement controlled variants of these circuits.  There are many constructions for these controlled gates~\cite{BCD95,Jones2013,Selinger2013}.  Here we anticipate that the cost of state preparation for the CISD state will be sub-dominant to the cost of the simulation.  This means that minimizing the number of qubits needed is an important design goal.  Let us define $\Lambda_m(G)$ to be the $m$-controlled version of the gate $G$.  Then the gate $\Lambda_m(H)$ can be implemented using two $\Lambda_{n-1}(X)$ gates, a $\Lambda_1(H)$ gate and an ancilla qubit,

\[
      \Qcircuit @R 1em @C 1.5em {
		 &\ctrl{2} 			&\qw		&\ctrl{2}	&\qw\\
	\vdots& 	& 		   	&          	&		\vdots&\\
	&\ctrl{1}	&\qw		&\ctrl{1}&\qw\\
	\lstick{0}&\targ &\ctrl{1}&\targ&\qw&\lstick{0}\\
	&\qw 	&\gate{H}&\qw&\qw
}				
    \]
Controlled $T^q$ gates can be performed similarly,
\[
      \Qcircuit @R 1em @C 1.5em {
		 &\ctrl{2} 			&\qw		&\ctrl{2}	&\qw\\
	\vdots& 	& 		   	&          	&		\vdots&\\
	&\ctrl{1}	&\qw		&\ctrl{1}&\qw\\
	&\ctrl{1} &\qw&\ctrl{1}&\qw&\\
	\lstick{0}&\targ	&\gate{T^q}&\targ&\qw&\lstick{0}
}				
    \]
The resulting circuits can be further optimized by noting that many of the Toffoli gates needed to perform the reductions of the least denominator exponent are redundant.  In particular we can express the simplified reduction circuit as,
\[
      \Qcircuit @R 1em @C 1.5em {
		 	&\ctrl{2} 		&\qw			&\qw			&\qw		&\qw		&\qw&&\ctrl{2}	&\qw\\
		\vdots& 			& 		   	&          		&		&		&~~~~\hdots&&	&\vdots 	&\\
			&\ctrl{1}		&\qw			&\qw			&\qw		&\qw		&\qw&&\ctrl{1}	&\qw\\
	\lstick{0}	&\targ 		&\ctrl{1}		&\ctrl{1}		&\qw		&\ctrl{1}	&\qw&&\targ		&\qw&\lstick{0}\\
			&\qw 			&\gate{H}		&\ctrl{1}		&\qw		&\ctrl{1}	&\qw~~~~\hdots&&\qw\\
	\lstick{0}	&\qw			&\qw			&\targ			&\gate{T^q}&\targ		&\qw&&\qw &
}				
    \]
The gate $\Lambda_1(H)$ requires $2$ $T$-gates~\cite{GS13}, and the Toffoli gates can be implemented, up to an irrelevant phase, using $4$ $T$-gates~\cite{Jones2013,Selinger2013} and an ancilla qubit.  The entire process requires at most $N+4$ qubits, which is typically less memory than is required for the quantum simulation and eigenvalue estimation phases of the algorithm.  This means that the additional four qubits required for the state preparation algorithm will not impact the memory requirements of the overall simulation algorithm.

For the present problem, the CISD state is in $\mathbb{C}^{2^{N +1}}$ (recall one additional qubit is needed to ensure a solution to the norm equations for synthesis).  This means that we also need to consider the cost of implementing $\Lambda_{N}(X)$ gates.
Although highly time-efficient constructions for the multiply controlled  circuits can be made using the circuits of~\cite{Jones2013}, they require a large number of qubits.  In order to ensure that the space complexity of state
preparation does not dominate the algorithm, we use the less time-efficient construction of Barenco et al~\cite{BCD95} to compile the $\Lambda_{N}(X)$ gates.
Using Corollary 7.4 from~\cite{BCD95} and the $\Lambda_{2}(iX)$ gate from~\cite{Jones2013,Selinger2013} to implement the Toffoli gate, the cost of implementing such circuits is at most
\begin{align}
T_{\rm count}\big(\Lambda_{N}(iX)\big)&\le 32(N-3).
\end{align}
At most $N+4$ qubits, where $N \ge 5$, are needed to implement these gates~\cite{BCD95}.  

The reduction of each of the two dimensional subspaces requires two steps.  First, the application of the $\Lambda_{N}(X)$ gates to mark the subspace and a sequence of $k$ controlled operations to reduce the denominator exponent of that subspace.  This reduction process requires $k$ steps, each of which involves at most two $\Lambda_1(H)$ and two $\Lambda_2(T^q)$.  Including the cost of the two $\Lambda_{N}(X)$ gates, the total cost of the reduction is at most $22k + 64(N-3)$ $T$-gates.
At most $D+1$ reduction steps are required in this process as well as potnentially a swap of the final state into the $\ket{0^{N+1}}$ state (which can be performed using Clifford operations).  Thus the overall $T$--count for this process is
\begin{equation}
(22k + 64(N-3))(D+1).
\end{equation}
This also is the $T$--count for for preparing the CISD state from the $\ket{0^{N+1}}$ state because the necessary circuit can be found by taking the Hermitian conjugate of the resultant gate sequence.
Thus using~\eq{keqn} the total number of non-Clifford operations required in the state preparation scales at most as

\begin{align}
\mathcal{O}\left({D}\log\left(\frac{D}{\delta}\right)+ND \right)
\end{align}

If the approach of Wang et al~\cite{WAF09},  coupled with recent methods for decomposing single qubit rotations into $T$ gates, is used to prepare the CISD state then the resultant $T$-count scales at most as $\tilde{\mathcal{O}}(2^{n_e} N^{n_e}/n_e!)\log(1/\delta)$.  If $n_e \approx N/2$ then this method is inefficient, whereas ours is not since $D\in\mathcal{O}(N^4)$ for CISD states.  If $n_e \le 3$ then the method of Wang et al does provide superior scaling as $N$ increases, though cases where $n_e \le 3$ and $N$ is large may be rare.  In contrast, the method of Ortiz et al~\cite{OGK01} requires $\tilde{\mathcal{O}}(D^2N^2 \log(1/\delta)$ gates, which is nearly quadratically slower than our method.

As a final point, the cost of the state preparation algorithm is $\mathcal{O}(N^5)$ in worst case scenarios.  This can be comparable to, or greater than, the cost of quantum simulation in the limit of large $N$.  This means that using a na\"ive CISD approximation in cases with half filling may seriously degrade the performance of the algorithm.  This means that in order to see the performance advantages promised by recent algorithms, which have scaling near $\mathcal{O}(N^4)$, sophisticated state preparation methods are needed in cases where the Hartree-Fock state has poor overlap with the FCI ground state.

\section{Conclusion}

Our work calls into question the basic assumption that the error in Trotter-Suzuki based methods for simulating quantum chemistry is \emph{explicitly} a function of the number of spin-orbitals used to represent the system.  We find through numerical evidence that such errors do not seem to be directly related to the number of spin-orbitals in the system for small molecules. We observe this lack of correlation for a variety of orbital bases including local, canonical and natural orbitals.  Instead, we see that chemical features such as the maximum nuclear charge is a strong indicator of the complexity of a simulation.  We argue that the errors should scale as $\mathcal{O}(Z_{\max}^6)$ for an atomic orbital basis, which is in close agreement with the scaling observed numerically.  We also observe that some atoms, such as Oxygen, Fluorine and Neon, have vanishingly small Trotter errors despite available error bounds predicting large Trotter errors for these molecules.  We show that this discrepancy can be understood as a consequence of the large filling fraction for these molecules.  This suggests that chemical features of a molecule may be much better predictors of the number of Trotter steps needed in a simulation than the number of spin-orbitals assigned to the molecule.

We further analyze the errors and see that the discrepancy between the observed Trotter error and the norm of the error operator does not arise from random cancellation.  Indeed, the errors observed are much greater than what would be expected if the ground state were a Haar random state that was chosen independently from the eigenvectors of the error operator.  Furthermore, we observe that the distribution of eigenvalues of the error operator is highly structured and has many near-zero eigenvalues, which likely is the cause of the orders of magnitude separation between the Trotter error and the norm of the error operator.

We also use the error operator to improve quantum simulation methods by providing a computationally efficient algorithm for estimating the error in a simulation.  This leads to two applications: ($a$) compensating for Trotter error in a quantum simulation by subtracting the prediction off the result and ($b$) predicting the number of Trotter steps needed in a simulation.  Finally, we provide a quantum algorithm for preparing CISD states that is polynomially more efficient than existing methods and may provide a viable alternative to adiabatic state preparation in cases where the Hartree-Fock approximation to the ground state leads to poor success probability.

There are several natural avenues of inquiry that this work reveals.  First, although this work shows strong numerical evidence for small molecules we do not have sufficient evidence to state conclusively that the error in the Trotter-Suzuki formula is independent of $N$ in the asymptotic limit.  Larger numerical experiments may be needed to shed more light on the scaling of Trotter--Suzuki errors in this regime.  Secondly, Ferredoxin is often suggested as a strong candidate for quantum chemistry simulation but Fe$_2$S$_2$ has large nuclear charges which make it a challenging molecule from the perspective of simulation.  This suggests that there may be other large organic molecules with smaller nuclear charges that may be even more natural targets for quantum simulation.  Finally, although our work has suggested that the number of spin-orbitals in a molecule may not uniquely characterize the cost of a quantum chemistry simulation, it does not provide a simple criteria for determining which molecules are easy or hard to simulate.  Finding molecular features, beyond the maximum nuclear charge and the filling fraction, that can be used to predict the relative difficulty of simulation would not only constitute an important step forward for quantum chemistry simulations but would also be an important contribution to quantum chemistry as a whole.

\section{Acknowledgements}
The authors thank Matthew Hastings and David Gosset for helpful conceptual discussions, Peter Love for comments on an early manuscript, John Parkhill for conversations regarding classical electronic structure methods and Vadym Kliuchnikov for discussions about preparation of the CISD state. J.M. is supported by the Department of Energy Computational Science Graduate Fellowship under grant number DE-FG02-97ER25308. A.A-G. appreciates the support of the Air Force Office of Scientific Research under award number FA9550-12-1-0046 and the National Science Foundation under award number CHE-1152291.

\bibliographystyle{apsrev4-1}
\bibliography{library,nathan}

\newpage
\appendix
\section{Computation of Haar Expectations}\label{app:Haar}
In order to determine whether the error cancellations observed for ground state quantum simulations arise because of properties of the eigenstates of the Hamiltonian, we need to determine whether these results would also be typical of random vectors.  Here we provide a derivation, for completeness, of the Haar expectation value and variance of the $|\braket{v}{k}|^2$. 

In the following we will take $k$ to be fixed and $v$ to represent the Haar random variable.
We will also use the convention that $\mathbb{E}_H$ denotes the expectation value of a quantity over a set of Haar random vectors, and $\mathbb{V}_H$ denotes the variance over the set.  To be clear, $\mathbb{E}_H |\braket{v}{k}|^2 = \int_{U \in {\rm Haar}} |\bra{0}U^\dagger\ket{k}|^2\mathrm{d}U$.

We wish to compute the variance, 
\begin{align}
\mathbb{V}_H (|\braket{v}{k}|^2) = \mathbb{E}_H(|\braket{v}{k}|^4) - \mathbb{E}_H(|\braket{v}{k}|^2)^2,
\end{align}
of the square of the overlap of an arbitrary Haar random vector, $\ket{0}$, with an eigenvector of an arbitrary Hermitian operator (in this case, the Trotter error operator), $\ket{k}$. We begin by stating the correspondence,
\begin{align}
\ket{v}\bra{v} = U \ket{0} \bra{0} U^\dagger
\end{align}
where $U$ is the unitary Gram matrix which affects a basis transformation into the error operator eigenbasis (for instance), and $\ket{v}$ represents $\ket{0}$ in the error operator eigenbasis. We are interested in the projection of this state onto an eigenvector of the error operator,
\begin{align}
a_k & = \langle k \ket{v} = \bra{k} U \ket{0}\\
| a_k |^2 & = \bra{k} U \ket{0}\bra{0} U^\dagger \ket{k}\\
 & = \textrm{tr}\left[\ket{k}\bra{k} U \ket{0}\bra{0} U^\dagger\right]\nonumber.
\end{align}
We compute this trace in two steps.  From the unitary invariance of the Haar measure we have that
\begin{align}
\int_{U(n)}\left[U \ket{0} \bra{0} U^\dagger\right]\mathrm{d}U= \frac{\openone}{2^N}
\end{align}
Therefore 
\begin{align}
 \mathbb{E}_H(| a_k |^2) = \textrm{tr}\left[\ket{k}\!\!\bra{k} \frac{\openone}{2^n}\right] = \frac{1}{2^N}.
\end{align}
Thus, $\mathbb{E}_H (|a_k|^2) = \frac{1}{2^N}$, and hence $\mathbb{E}_H(|a_k|^2)^2 = \frac{1}{2^{2 N}}$.

Focusing on the remaining component of the variance,
\begin{align}
| a_k |^4 & = \bra{k} U \ket{0}\bra{0} U^\dagger \ket{k}\bra{k} U \ket{0}\bra{0} U^\dagger \ket{k}\\
 & = \textrm{tr}\left[\left(\ket{k} \bra{k}\right)^{\otimes 2} U^{\otimes 2} \left(\ket{0}\bra{0}\right)^{\otimes 2} U^{\dagger \otimes 2} \right].\nonumber
\end{align}
To further evaluate the trace we follow the treatment in \cite{hayden2006aspects} which uses the spectral theorem to derive orthogonal projectors onto symmetric and antisymmetric subspaces. This begins by defining a flip operator, $\mathbb{F} \in \mathbb{C}^{2^{2 N} \times 2^{2 N}}$,
\begin{align}
\mathbb{F} \left(\ket{\psi} \otimes \ket{\varphi}\right) = \ket{\varphi} \otimes \ket{\psi}.
\end{align}
From this definition it is clear that
\begin{align}
\mathbb{F} = \pi_\textrm{sym} - \pi_\textrm{antisym}\\
\openone^{\otimes 2} =  \pi_\textrm{sym} + \pi_\textrm{antisym}.
\end{align}
Thus,
\begin{align}
\pi_\textrm{sym} = \frac{1}{2}\left(\openone^{\otimes 2} + \mathbb{F}\right)\\
\pi_\textrm{antisym} = \frac{1}{2}\left(\openone^{\otimes 2} - \mathbb{F}\right).
\end{align}
Since $\textrm{tr}\left[\openone^{\otimes 2}\right] = 2^{2 N}$ and $\textrm{tr}(\mathbb{F})=2^N$,
\begin{align}
\textrm{tr}\left[\pi_\textrm{sym}\right] = \frac{2^N \left(2 ^N + 1\right)}{2}\\
\textrm{tr}\left[\pi_\textrm{sym}\right] = \frac{2^N \left(2 ^N - 1\right)}{2}.
\end{align}
Since $\ket{0}\bra{0} \otimes \ket{0}\bra{0}$ is entirely symmetric, it is straight forward to see from unitary invariance that
\begin{align}
\int_{U(n)}\left[U^{\otimes 2}\left( \ket{0}\bra{0}\right)^{\otimes 2} U^{\dagger \otimes 2}\right]\mathrm{d}U = \frac{2}{2^N \left(2^N + 1\right)} \pi_\textrm{sym}.
\end{align}
Thus,
\begin{align}
\mathbb{E}_H(| a_k |^4) &=  \frac{2}{2^N \left(2^N + 1\right)} \textrm{tr}\left[\pi_\textrm{sym} \left(\ket{k}\bra{k}\right)^{\otimes 2}\right]\nonumber\\
& = \frac{2}{2^N \left(2^N + 1\right)}.
\end{align}
Finally, we arrive at the variance of $| a_k |^2$,
\begin{align}
\mathbb{V}_H(|a_k|^2) =  \frac{2}{2^N \left(2^N + 1\right)} - \frac{1}{2^{2 N}}.
\end{align}

This gives us the variance in the $|a_k|^2$ terms, which in turn allows us to find the deviation from the expected error for a quantum chemistry simulation.  The key point here is that the standard deviation is on the order of the expectation value which means that we expect relatively large fluctuations in the probabilities that correspond to particular eigenvalues of the Trotter error operator.  Hence we do not expect a concentration of measure result to hold in high dimensional spaces.

\clearpage
\FloatBarrier
\onecolumngrid
\section{Contributions of Orbitals to Trotter Error Operator}\label{app:extradata}
In Section II we provided evidence that transitions involving the inner-most electrons contribute most to the error in the quantum simulation.  This is perhaps surprising given that transitions involving the valence electrons, rather than the core electrons,  are typically more relevant for understanding the properties of a molecule.  We present additional numerical results in Figure 11 that examine this for water and Beryllium hydride.  These results confirm the intuition developed earlier that interactions involving the two inner--most orbitals significantly impact the errors in the ground state energy.

We also observe a rough correlations between the magnitudes of the error coefficients in the expansion of the error operator and their contribution to the ground state error in the local orbital basis.  This suggests that looking at the contribution of the inner orbitals is the most significant for the error for these molecules and that the aggregate contributions of the error coefficients correlates roughly with the ground state error.

The analogous data for natural orbital basis in Figure 11 defies this approximate correspondence for both water and Beryllium hydride.  For the case of water, interactions that involve orbitals $7$ and $8$ are the second-largest contributors to  the error in the ground state energy.  The significance of these transitions is not apparent in the corresponding plots of the magnitude of the Hamiltonian coefficients nor the magnitudes of the error coefficients in the expansion of the error operator.   Similarly, for Berylium Hydride, interactions involving orbitals $7$ and $8$ may be expected to have a significant impact on the error in the ground-state energy but the data suggests that they do not.   These results underscore the challenges faced when attempting to understand the nature of the error operator from solely looking at the magnitudes of the error coefficients.
\vspace{-.5cm}
\begin{figure*}[!ht]
         \centering
         \subfloat[][Water, local basis, Hamiltonian coefficients.]{
                 \includegraphics[width=.3\textwidth]{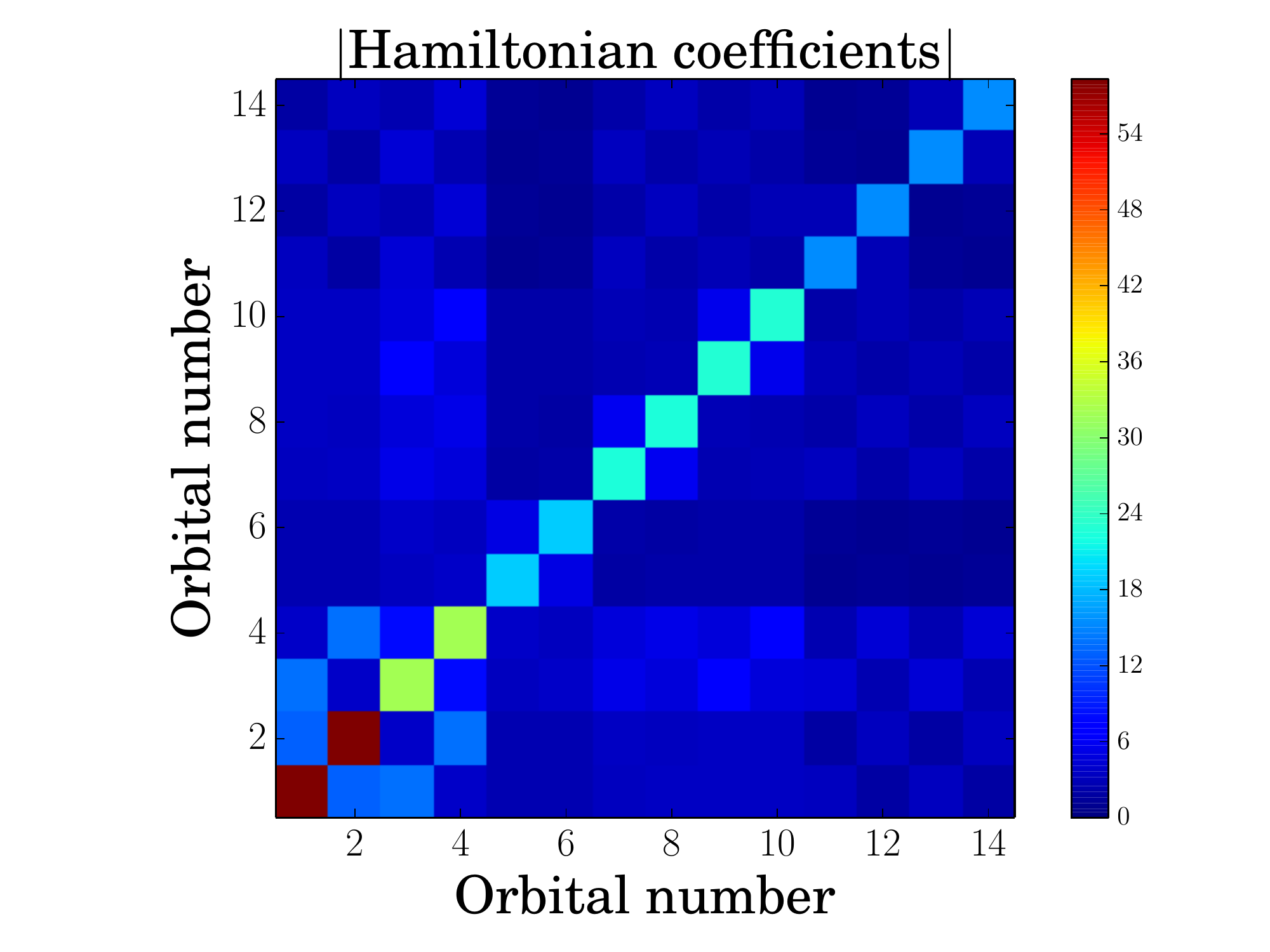}}
         \subfloat[][Water, local basis, error coefficients.]{
                 \includegraphics[width=.3\textwidth]{H2O_OAO_coefficients_double_orbitals.pdf}}
         \subfloat[][Water, local basis, error contributions.]{
                 \includegraphics[width=.3\textwidth]{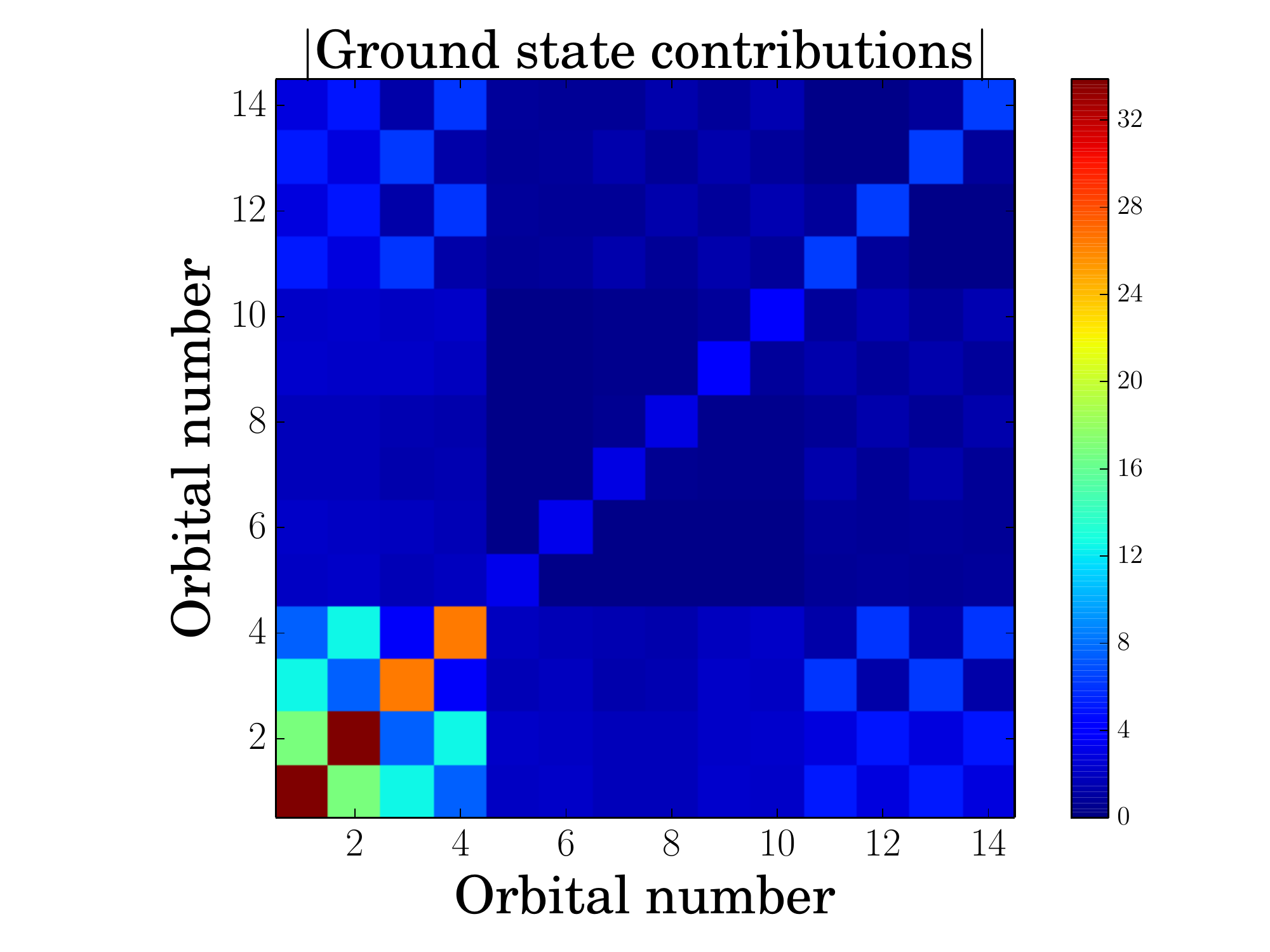}}\\
         \subfloat[][Water, natural basis, Hamiltonian coefficients.]{
                 \includegraphics[width=.3\textwidth]{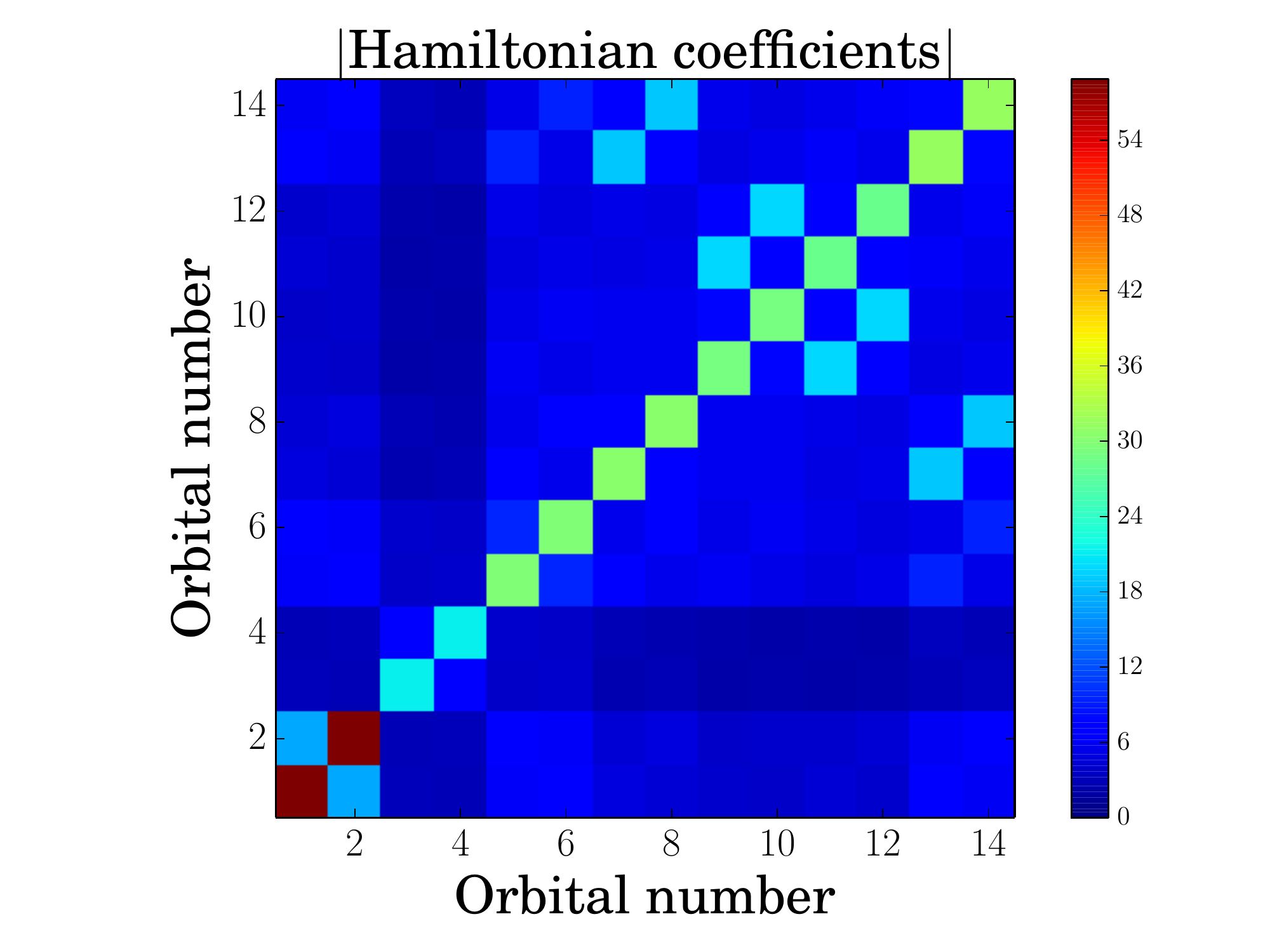}}
         \subfloat[][Water, natural basis, error coefficients.]{
                 \includegraphics[width=.3\textwidth]{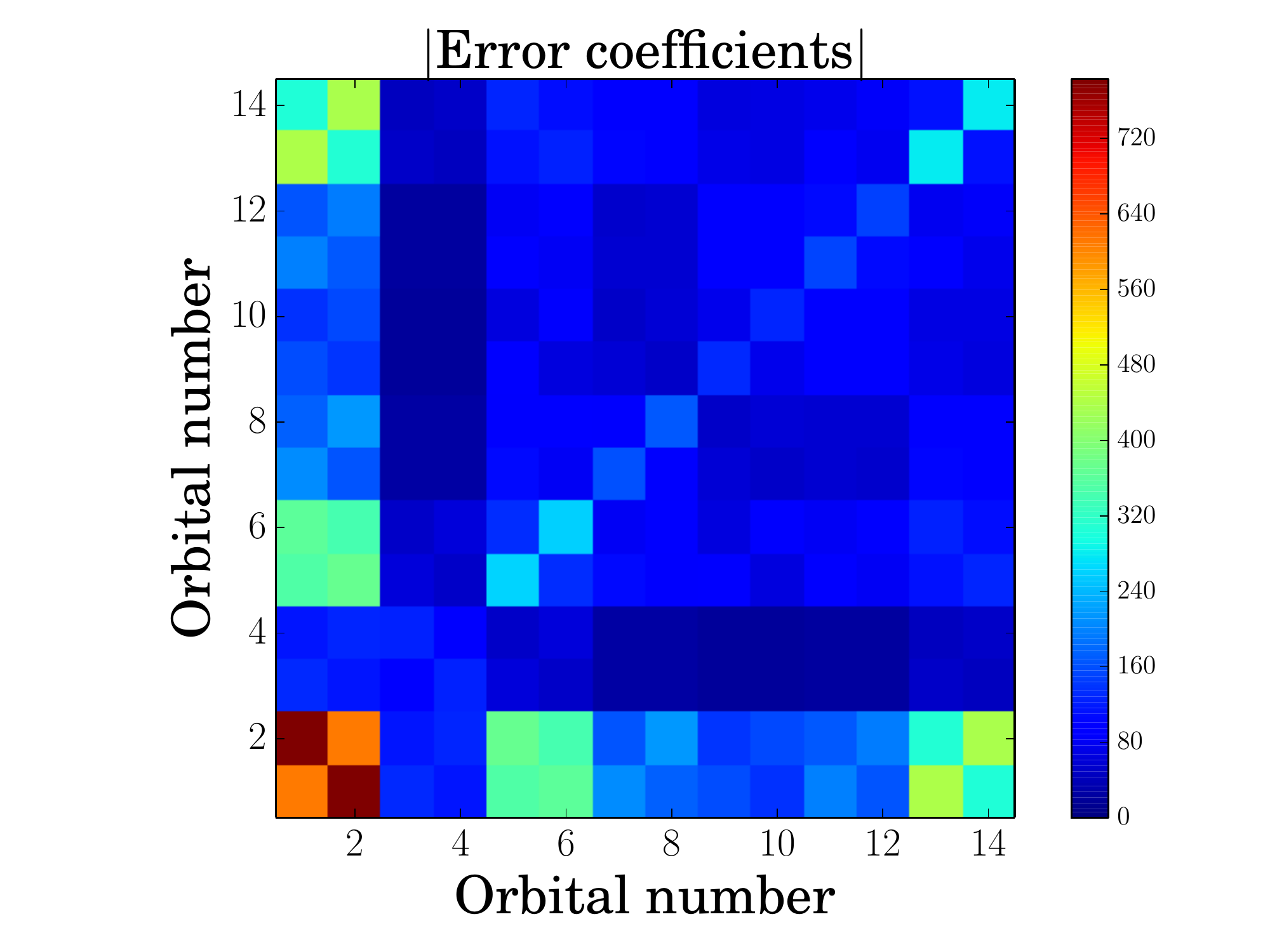}}
         \subfloat[][Water, natural basis, error contributions.]{
                 \includegraphics[width=.3\textwidth]{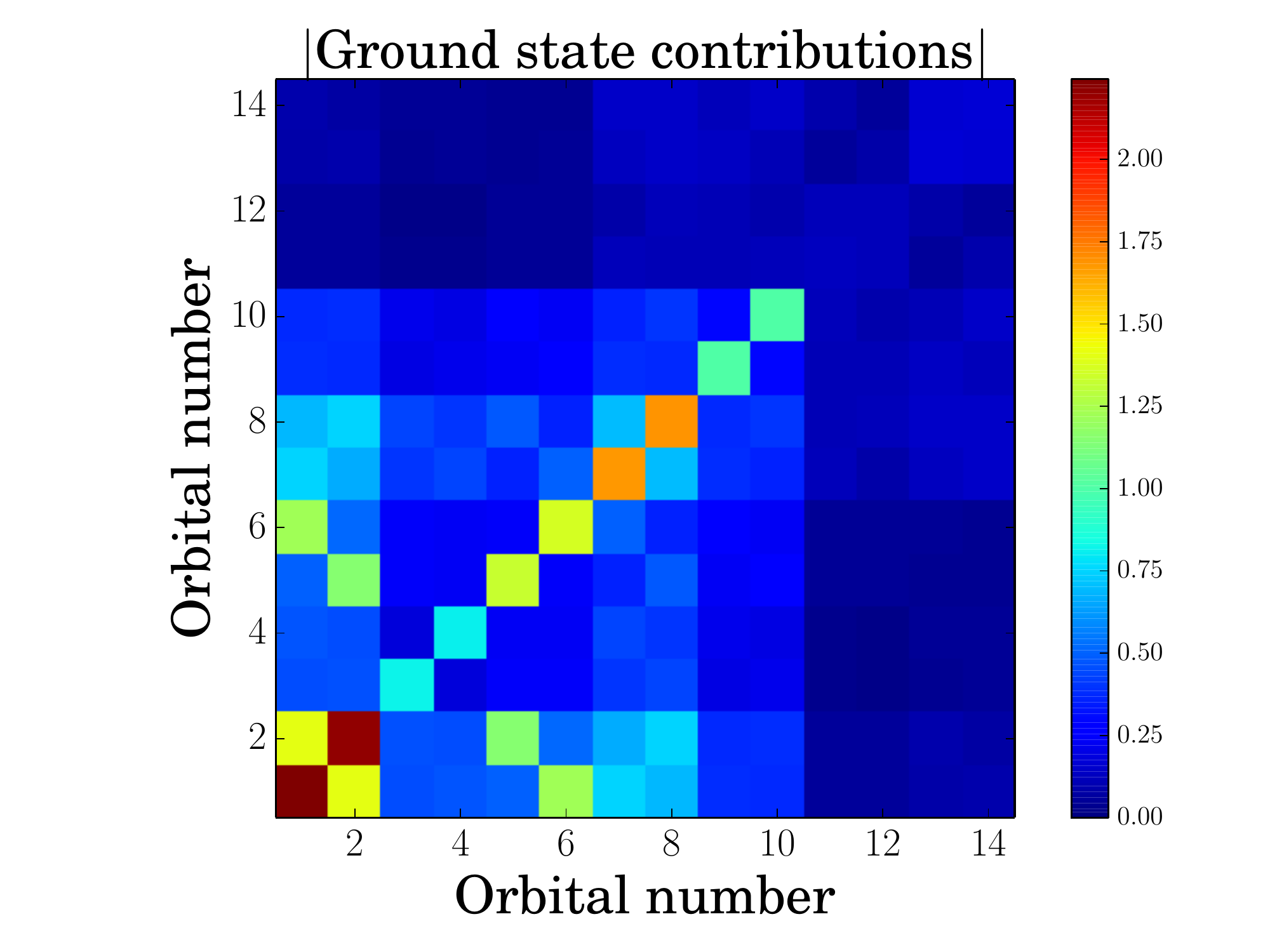}}\\
         \subfloat[][Beryllium hydride, local basis, Hamiltonian coefficients.]{
                 \includegraphics[width=.3\textwidth]{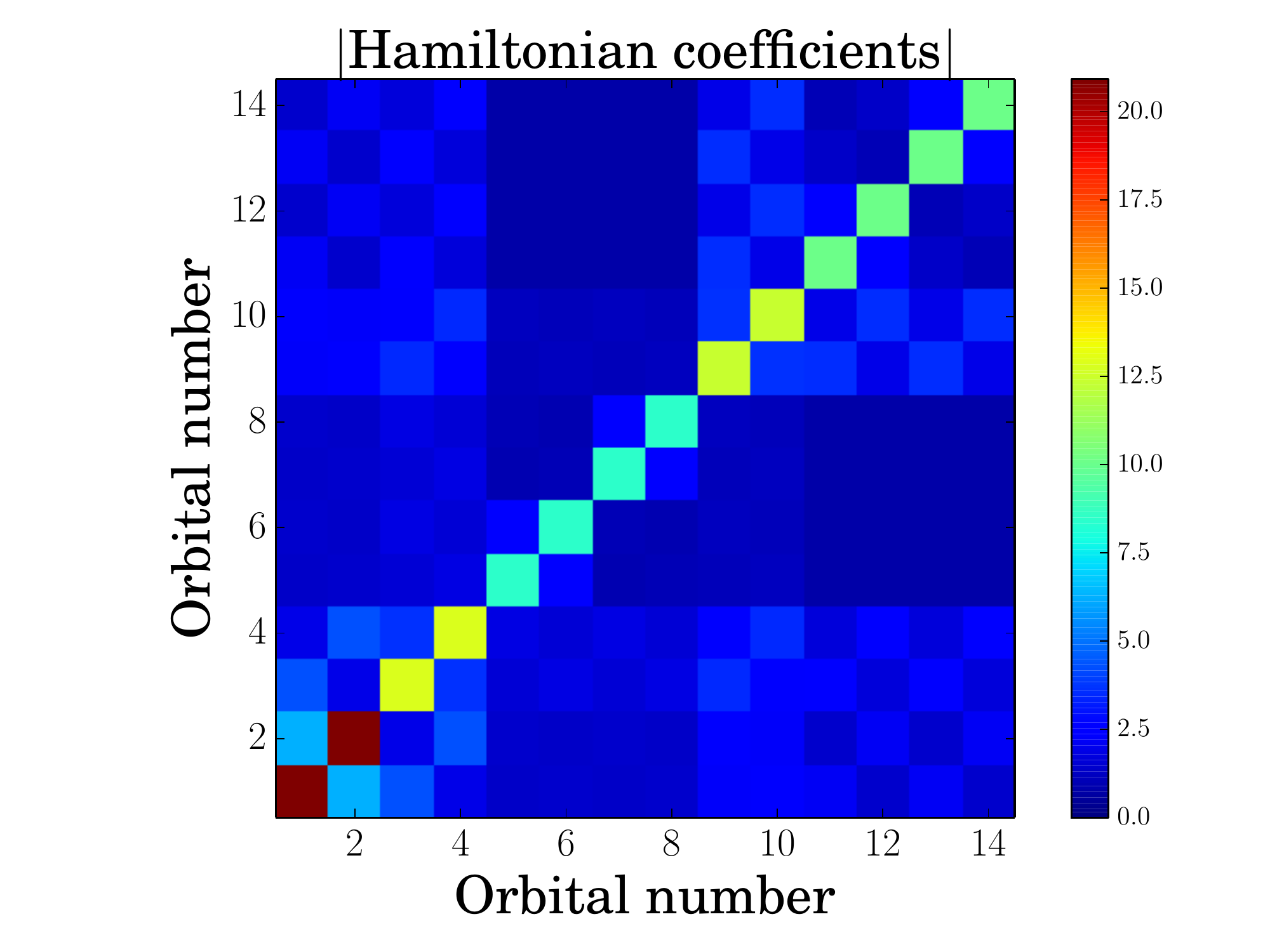}}
         \subfloat[][Beryllium hydride, local basis, error coefficients.]{
                 \includegraphics[width=.3\textwidth]{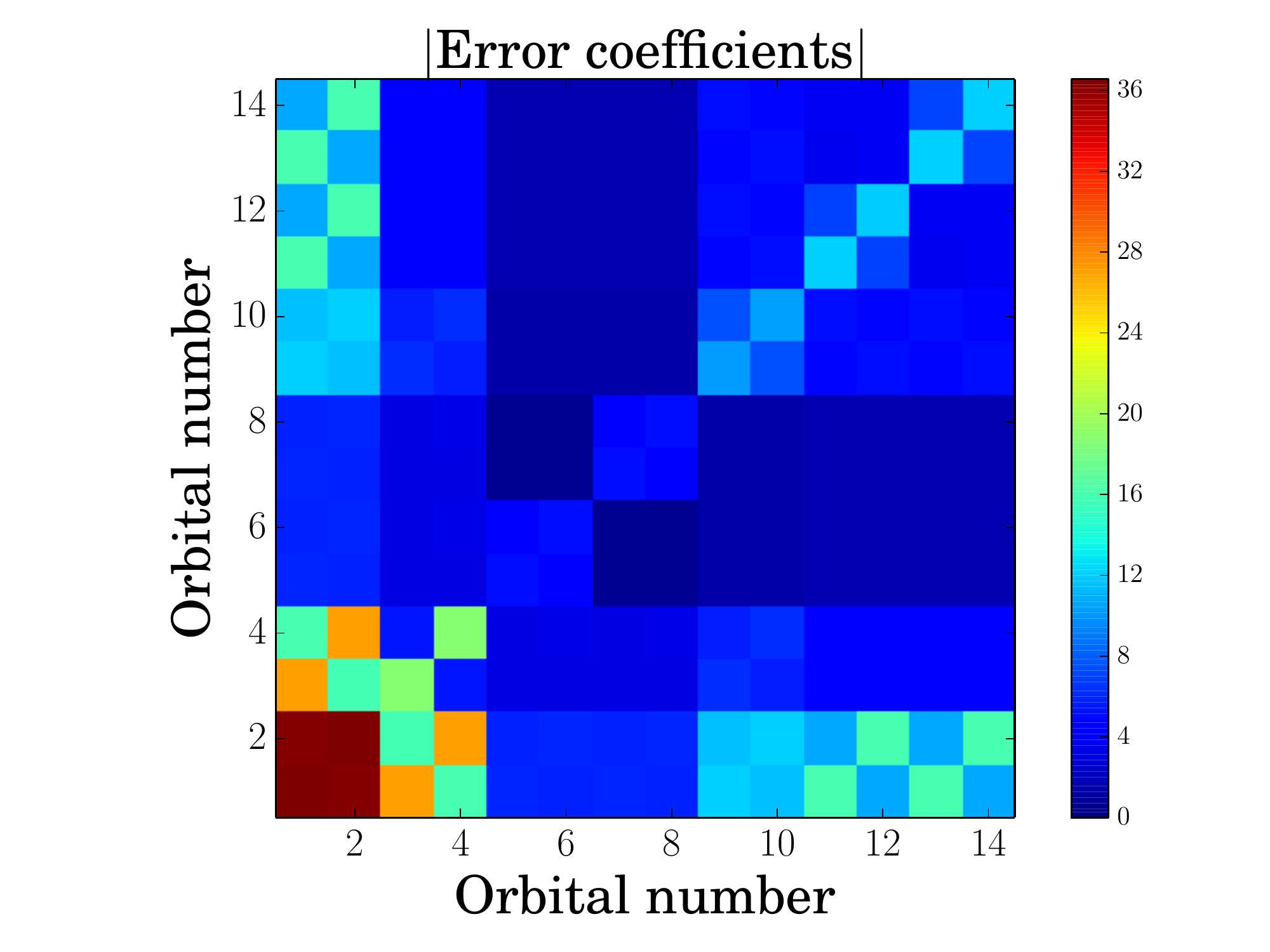}}
         \subfloat[][Beryllium hydride, local basis, error contributions.]{
                 \includegraphics[width=.3\textwidth]{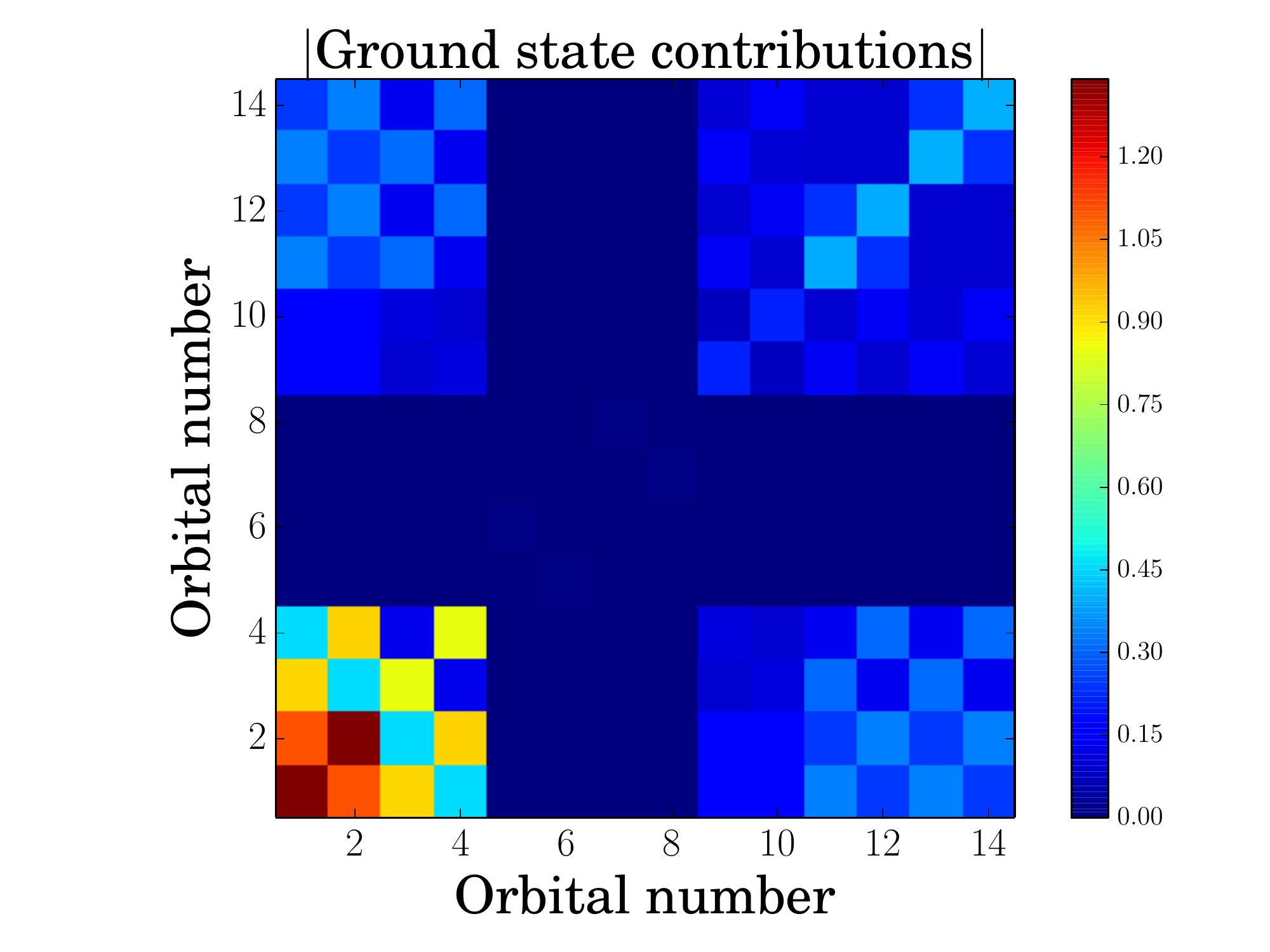}}\\
         \subfloat[][Beryllium hydride, natural basis, Hamiltonian coefficients.]{
                 \includegraphics[width=.3\textwidth]{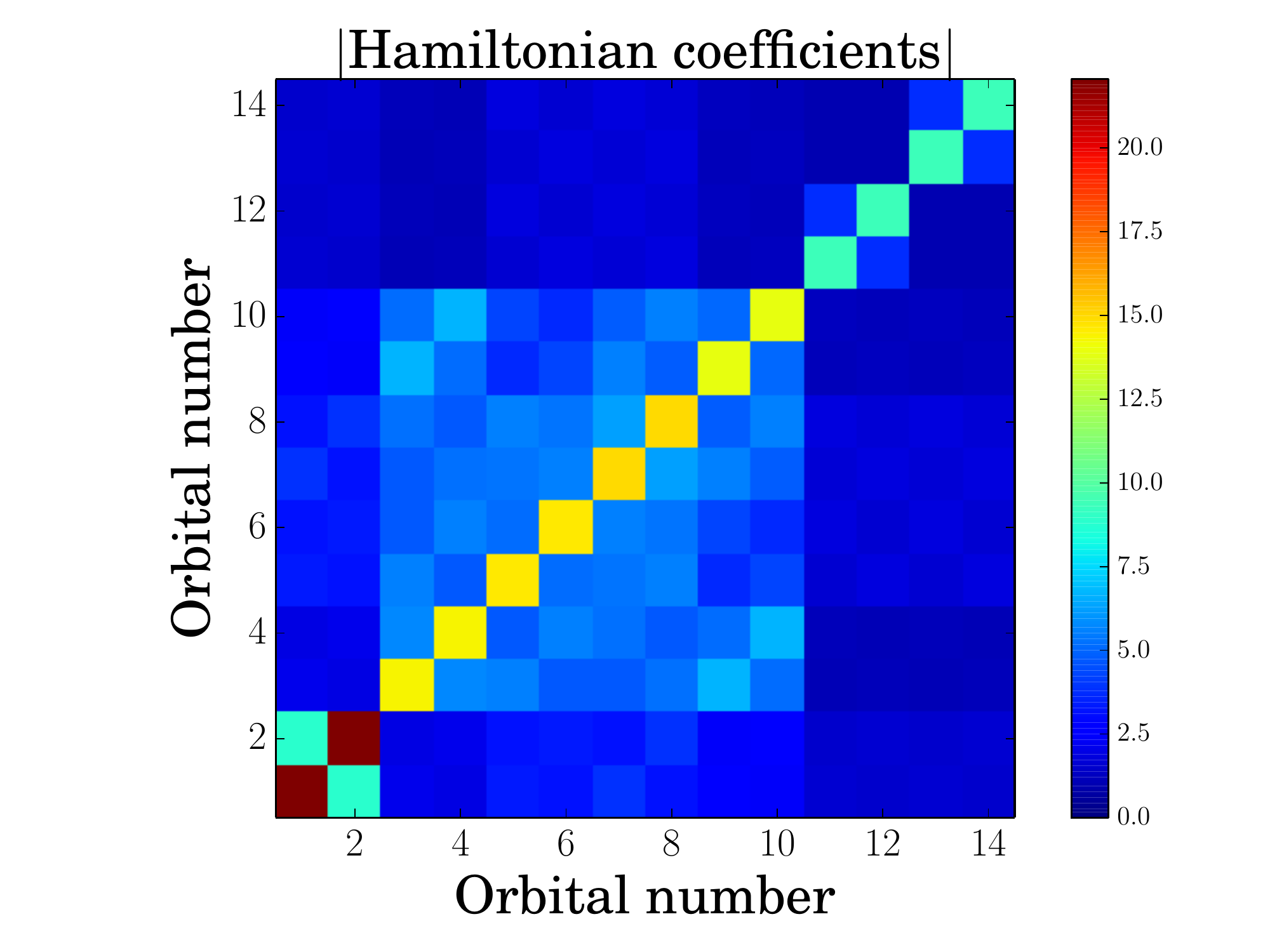}}
         \subfloat[][Beryllium hydride, natural basis, error coefficients.]{
                 \includegraphics[width=.3\textwidth]{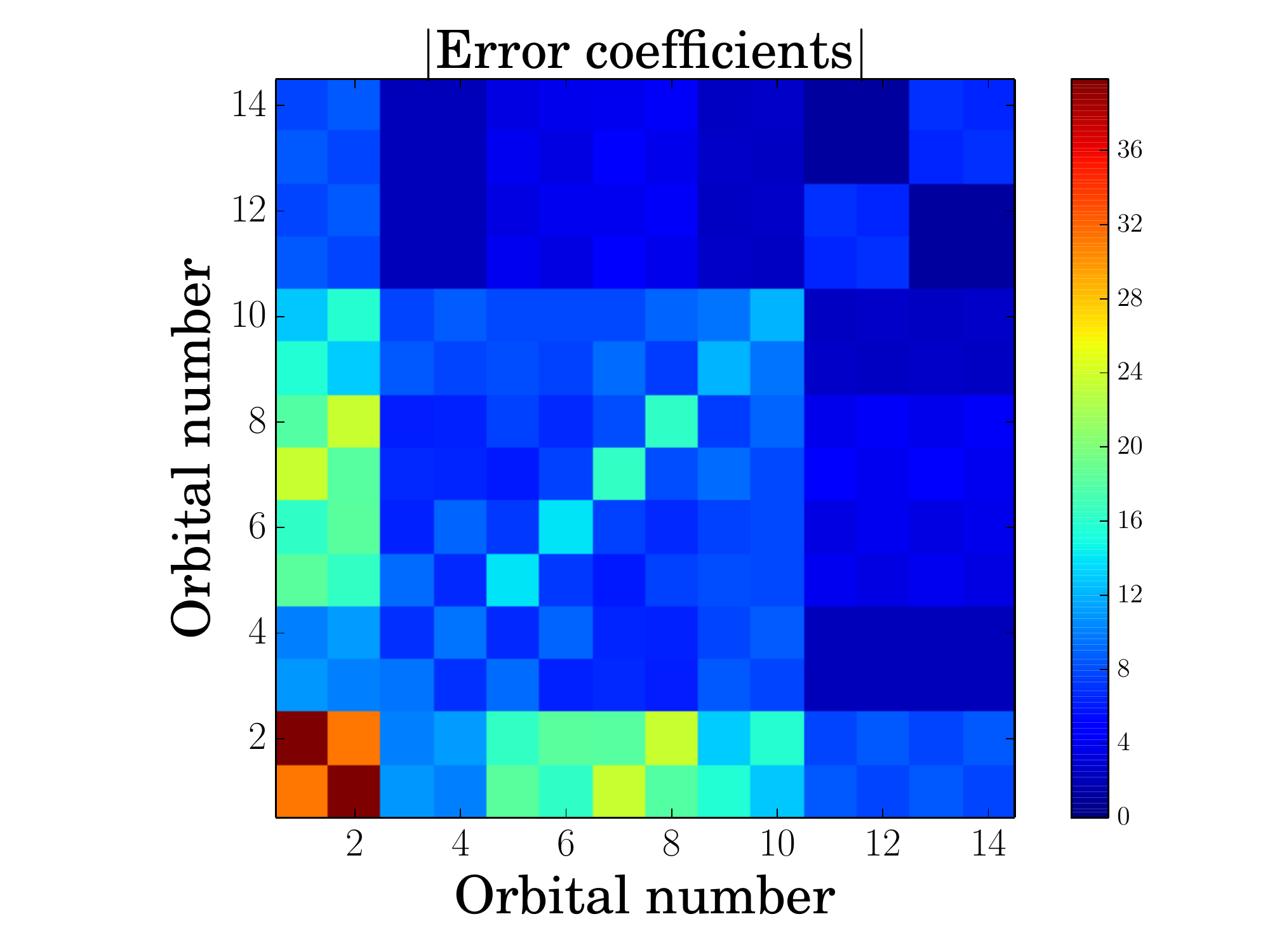}}
         \subfloat[][Beryllium hydride, natural basis, error contributions.]{
                 \includegraphics[width=.3\textwidth]{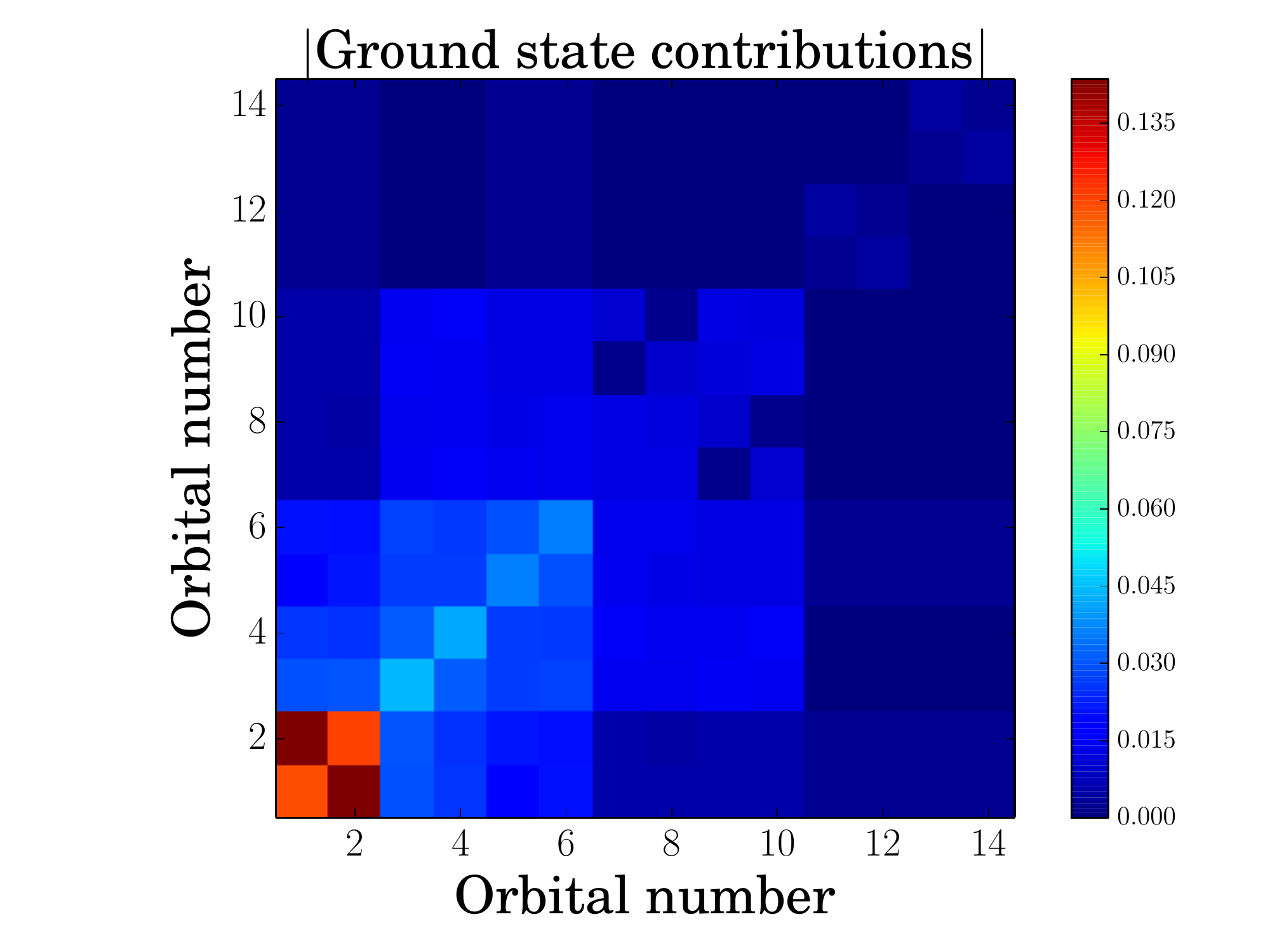}}\\
         \caption{These plots show the coefficients of normal-ordered terms in the Hamiltonian and error operator as well as expectation values of the error operator terms for the ground state. The terms are binned according to the orbitals involved in the term. This plot shows the marginal distribution of the magnitudes of these terms.\label{fig:moredata}}
\end{figure*}
\FloatBarrier

\end{document}